\documentclass[11pt]{article}
\pdfoutput=1

\usepackage{aas_macros,amsmath,amssymb,cite,esint,graphicx}
\usepackage[margin=.8in,letterpaper]{geometry}
\usepackage[colorlinks=true]{hyperref}
\usepackage[affil-it]{authblk}

\numberwithin{equation}{section}
\let\originalleft\left
\let\originalright\right
\renewcommand{\left}{\mathopen{}\mathclose\bgroup\originalleft}
\renewcommand{\right}{\aftergroup\egroup\originalright}
\mathcode`\*="8000
{\catcode`\*=\active\gdef*{\mathclose{}\,\mathopen{}}}

\newcommand{\ab}[1]{\left|#1\right|}
\newcommand{\br}[1]{\left[#1\right]}
\newcommand{\cu}[1]{\left\{#1\right\}}
\newcommand{\pa}[1]{\left(#1\right)}

\newcommand{\ed}{\mathop{}\!\mathrm{d}}
\newcommand{\pd}{\mathop{}\!\partial}
\renewcommand{\O}[1]{\mathcal{O}\pa{#1}}
\renewcommand{\mod}{\mathrm{mod}}

\begin{document}

\title{\Huge{Observational Signature of High Spin\\
at the Event Horizon Telescope}}
\date{}
\author[1]{Samuel E. Gralla\thanks{sgralla@email.arizona.edu}}
\author[2,3]{Alexandru Lupsasca\thanks{lupsasca@fas.harvard.edu}}
\author[2]{Andrew Strominger\thanks{strominger@physics.harvard.edu}}
\affil[1]{\small
Department of Physics, University of Arizona\\
Tucson, AZ 85721, USA}
\affil[2]{\small
Center for the Fundamental Laws of Nature, Harvard University\\
Cambridge, MA 02138, USA}
\affil[3]{\small
Society of Fellows, Harvard University\\
Cambridge, MA 02138, USA}

\maketitle

\begin{abstract}
	We analytically compute the observational appearance of an isotropically emitting point source on a circular, equatorial orbit near the horizon of a rapidly spinning black hole.  The primary image moves on a \textit{vertical} line segment, in contrast to the primarily horizontal motion of the spinless case.  Secondary images, also on the vertical line, display a rich caustic structure.  If detected, this unique signature could serve as a ``smoking gun'' for a high-spin black hole in nature.
\end{abstract}

\vfill\pagebreak

\tableofcontents

\newpage

\section{Introduction}

Over the last several decades, abundant astronomical evidence for black holes has accumulated from a variety of sources \cite{Narayan2013}, most notably the recent spectacular observations \cite{Abbott2016a,Abbott2016b,Abbott2017a,Abbott2017b} of gravitational waves emitted from black hole mergers.  In all of these observations, the black holes appear as point-like objects, as the detectors have been far from being able to resolve distances on the scale of the Schwarzschild radius.  The existence of the most striking feature of a black hole---namely, the  event horizon---is only indirectly inferred.

All of this is expected to change dramatically within the coming year, when the Event Horizon Telescope (EHT) obtains images of black holes comprised of pixels smaller than the Schwarzschild radius.

This opens an exciting new chapter in experimental black hole astrophysics.  It also presents a host of challenges to theorists who need to predict what will be seen by the EHT \cite{Bardeen1973,Luminet1979,Broderick2006,Doeleman2008a,Doeleman2008b,Doeleman2009,Doeleman2012,Johnson2015}.  While the Kerr solution is itself relatively simple, the nearby environment can contain complex magnetospheres, accretion disks, and jets that are the origin of the actual observed signal.  The predicted signal in general depends on many a priori undetermined parameters describing this environment.

Universal and sometimes striking predictions are possible for the case of rapidly spinning black holes \cite{Andersson2000,Glampedakis2001,Yang2013,Porfyriadis2014a,Porfyriadis2014b,Hadar2015,Gralla2016a,Gralla2016b,Burko2016,Hadar2017,Compere2017}.  At the maximal allowed value of the spin, $J=GM^2$, the region near the horizon of the black hole acquires an infinite-dimensional conformal symmetry \cite{Bardeen1973,Bardeen1999,Guica2009}.  This is a precise astrophysical analog of the universal critical behavior appearing in many condensed matter systems.  Not only do the symmetries supply powerful computational tools, but the universality reduces the dependence of physical predictions on undetermined parameters.  For example, it was found recently that gravitational waves from a near-horizon orbiting body can end with a slow decay to silence on a single characteristic frequency \cite{Gralla2016b}, in stark contrast to the rapid ``chirp'' of ordinary black hole binaries.

In this paper, we analyze the signal produced by a ``hotspot'' (localized emissivity enhancement) orbiting near a high-spin black hole, and find a very striking signal.  The primary image moves along a line segment (the  ``NHEKline"), which is rotated by $90^\circ$ relative to the orbital plane, just inside the shadow from black hole backlighting.  Secondary images are generally negligible except for bright caustic flashes which extend to the whole line segment.  These emissions pulsate in a complex periodic manner.  This signature is strongest in the edge-on case, when the observer lies near the equatorial plane ($\theta_o\approx90^\circ$), and disappears entirely when $\theta_o<\arctan{\pa{4/3}^{1/4}}\approx47^\circ$, a critical angle determined by near-horizon physics.  In general, emission signals of this type can only be computed numerically, but the emergent symmetries at high spin enable us herein to study the problem analytically.  We perform a detailed calculation for a uniformly emitting sphere orbiting in the equatorial plane, but the main conclusions generalize to all near-horizon sources.

Of course, it would require a fortuitous alignment of circumstances for an EHT target to have both high spin and a sufficiently long-lived brightness enhancement in the near-horizon region.  Nevertheless, as we move into the era of precision black hole observation, it is not unreasonable to hope that such a configuration might eventually be observed.  With enough resolution, the unique features of the signal would serve as a ``smoking gun'' for a high-spin black hole.

The paper is organized as follows.  In Sec.~\ref{sec:OrbitingEmitter}, we define the problem at any finite, non-extremal value of spin, and write down the equations to be solved.  In Sec.~\ref{sec:NearExtremalExpansion}, we solve the equations in the high-spin limit, keeping leading and subleading terms.  In Sec.~\ref{sec:ObservationalAppearance}, we explore the detailed observational appearance.  In App.~\ref{app:Shadow}, we discuss the black hole shadow in the high-spin regime.  In App.~\ref{app:NHEK}, we explain the connection of our computation to near-horizon geometry and argue that the signal persists more generally.  We relegate technical aspects of our calculations to the remaining appendices.

\begin{figure}[ht!]
	\includegraphics[width=\textwidth]{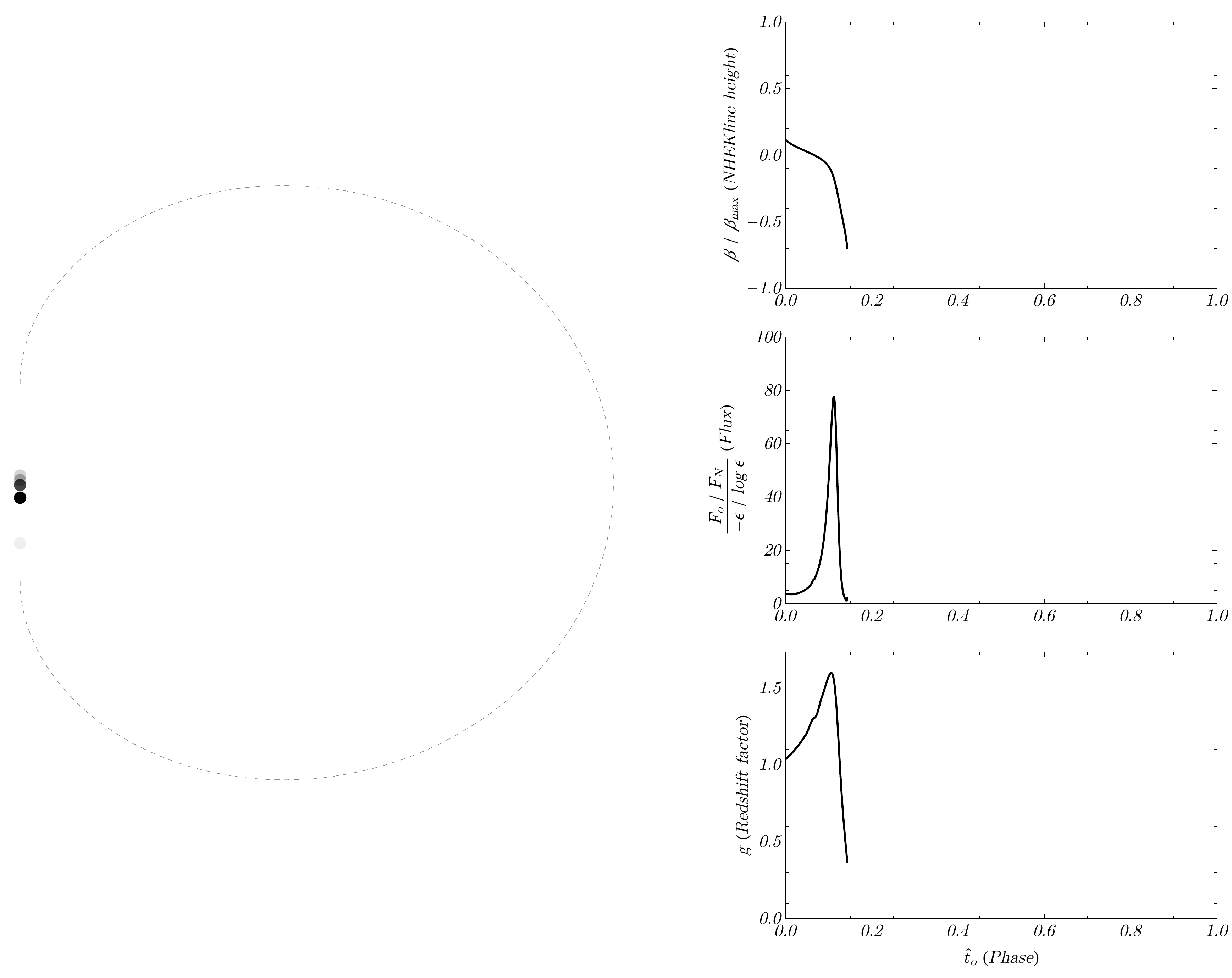}
	\caption{Observational appearance of a point emitter (``hotspot") orbiting near a rapidly spinning black hole.  All light appears on a vertical line segment, the so-called NHEKline, which forms a portion of the black hole shadow's edge (dashed line).  Over each cycle of the periodic image, the primary image appears near the center of the NHEKline before moving downward while blueshifting and spiking in brightness (right panels).  On the right, we display the height of the image on the NHEKline (relative to the maximum), its flux (relative to a comparable Newtonian problem, with spin-dependence $\epsilon$ of the black hole scaled out), and its redshift factor (ratio of observed to emitted frequency).  Notice the net blueshift ($g>1$) at peak brightness, reflecting the Doppler boost from the ultrarelativistic near-horizon orbit overcoming the gravitational redshift.	 Video animations are available \href{https://youtu.be/ZRfTj5JKPkA}{here}.  Secondary images have a rich caustic structure shown in Fig.~\ref{fig:3x3} below.  (The primary image, depicted here in black, is colored green in Fig.~\ref{fig:3x3}.)  The position of the source at the time of emission is shown in Fig.~\ref{fig:Winding} for the primary image.  The spin is $a/M=99.99995\%$ ($\epsilon=.01$) and the viewing angle is nearly edge-on, $\theta_o=84^\circ$.  Complete parameter choices are given in  Eq.~\eqref{eq:ExampleParameters}.  The appearance is qualitatively similar for other parameter choices.}
	\label{fig:OpticalAppearance}
\end{figure}

\section{Orbiting emitter}
\label{sec:OrbitingEmitter}

We work in the Kerr spacetime in Boyer-Lindquist coordinates $(t,r,\theta,\phi)$.  The metric is
\begin{align}
	ds^2=-\frac{\Delta\Sigma}{\Xi}\ed t^2+\frac{\Sigma}{\Delta}\ed r^2+\Sigma\ed\theta^2+\frac{\Xi\sin^2{\theta}}{\Sigma}\pa{\ed\phi-\omega\ed t}^2,
\end{align}
where
\begin{align}
	\omega=\frac{2aMr}{\Xi},\qquad
	\Delta=r^2-2Mr+a^2,\qquad
	\Sigma=r^2+a^2\cos^2{\theta},\qquad
	\Xi=\pa{r^2+a^2}^2-\Delta a^2\sin^2{\theta}.
\end{align}
Our emitter will be a point source orbiting on a circular, equatorial geodesic at radius $r_s$.  The angular velocity is \cite{Bardeen1972}
\begin{align}
	\Omega_s=\pm\frac{M^{1/2}}{r_s^{3/2}\pm aM^{1/2}},
\end{align}
where the upper/lower sign corresponds to prograde/retrograde orbits.  Here and hereafter, the subscript $s$ stands for ``source."  The local rest frame of the emitter consists of the four-velocity $u^\mu=e_{(t)}^\mu$ ($u_\mu u^\mu=-1$) along with three orthogonal unit spacelike vectors,
\begin{subequations}
\label{eq:EmitterFrame}
\begin{align}
	e_{(t)}&=\gamma\sqrt{\frac{\Xi}{\Delta\Sigma}}\pa{\pd_t+\Omega_s\pd_\phi},\qquad
	e_{(r)}=\sqrt{\frac{\Delta}{\Sigma}}\pd_r,\qquad
	e_{(\theta)}=\frac{1}{\sqrt{\Sigma}}\pd_\theta,\\
	e_{(\phi)}&=\gamma v_s\sqrt{\frac{\Xi}{\Delta\Sigma}}\pa{\pd_t+\omega\pd_\phi}+\gamma\sqrt{\frac{\Sigma}{\Xi}}\pd_\phi,
\end{align}
\end{subequations}
where\footnote{Note that $v_s$ and $\gamma$ are the velocity and boost factor according to the zero-angular momentum observer with four-velocity proportional to $\pd_t+\omega\pd_\phi$.}
\begin{align}
	\label{eq:EmitterVelocity}
	v_s=\frac{\Xi}{\Sigma\sqrt{\Delta}}\pa{\Omega_s-\omega}
	=\frac{\pm M^{1/2}\pa{r_s^2\mp2aM^{1/2}r_s^{1/2}+a^2}}{\sqrt{\Delta}\pa{r_s^{3/2}\pm aM^{1/2}}},\qquad
	\gamma=\frac{1}{\sqrt{1-v_s^2}}.
\end{align}
We define frame components of four-vectors $V^\mu$ in the usual way,
\begin{align}
	V^{(b)}=\eta^{(a)(b)}e_{(b)}^\mu V_\mu,
\end{align}
where $\eta^{(a)(b)}=\mathrm{diag}\pa{-1,1,1,1}$ and summation over repeated indices is implied.  We raise and lower frame indices with $\eta^{(a)(b)}$.

\subsection{Photon conserved quantities and interpretation} 

The wavelength of light from astrophysically realistic sources is much smaller than the size of the black hole.  This allows us to work in the geometric optics limit, where the emission corresponds to photons traveling on null geodesics.  Each such photon with four-momentum $p^\mu$ possesses four conserved quantities:
\begin{enumerate}
\item
the invariant mass $p_\mu p^\mu=0$,
\item
the total energy $E=-p_t$,
\item
the component of angular momentum parallel to the axis of symmetry $L=p_\phi$, and
\item
the Carter constant $Q=p_\theta^2-\cos^2{\theta}\pa{a^2p_t^2-p_\phi^2\csc^2{\theta}}$.
\end{enumerate}
The trajectory of the photon is independent of its energy and may be described by two rescaled quantities,
\begin{align}
	\label{eq:ConservedQuantities}
	\hat{\lambda}=\frac{L}{E},\qquad
	\hat{q}=\frac{\sqrt{Q}}{E}.
\end{align}
We follow the conventions of Refs.~\cite{Cunningham1972,Cunningham1973}, but put hats over these quantities to distinguish them from the unhatted $\lambda$ and $q$ that we introduce in Sec.~\ref{sec:NearExtremalExpansion}.  Note that while $Q$ can be negative and therefore $\hat{q}$ imaginary, only $\hat{q}^2$ appears in subsequent formulas.  Furthermore, since $Q=p_\theta^2\ge0$ when $\theta=\pi/2$, any photon passing through the equatorial plane must have a nonnegative Carter constant, and hence real $\hat{q}$.  Since we restrict to photons emitted by an equatorial source, we will always have real $\hat{q}>0$.  

The four-momentum may be reconstructed from the conserved quantities up to two choices of sign corresponding to the direction of travel,
\begin{subequations}
\label{eq:GeodesicEquation}
\begin{align}
	\label{eq:RadialGeodesicEquation}
	\frac{\Sigma}{E}p^r&=\pm\sqrt{\mathcal{R}(r)},\\
	\label{eq:PolarGeodesicEquation}
	\frac{\Sigma}{E}p^\theta&=\pm\sqrt{\Theta(\theta)},\\
	\label{eq:AzimuthalGeodesicEquation}
	\frac{\Sigma}{E}p^\phi&=-\pa{a-\frac{\hat{\lambda}}{\sin^2{\theta}}}+\frac{a}{\Delta}\pa{r^2+a^2-a\hat{\lambda}},\\
	\frac{\Sigma}{E}p^t&=-a\pa{a\sin^2{\theta}-\hat{\lambda}}+\frac{r^2+a^2}{\Delta}\pa{r^2+a^2-a\hat{\lambda}},
\end{align}
\end{subequations}
where 
\begin{subequations}
\begin{align}
	\mathcal{R}(r)&=\pa{r^2+a^2-a\hat{\lambda}}^2-\Delta\br{\hat{q}^2+\pa{a-\hat{\lambda}}^2},\\
	\Theta(\theta)&=\hat{q}^2+a^2\cos^2{\theta}-\hat{\lambda}^2\cot^2{\theta}.
\end{align}
\end{subequations}
The functions $\mathcal{R}(r)$ and $\Theta(\theta)$ are generally called the radial and angular ``potentials''.  Zeros of these functions correspond to turning points in the trajectories, where the sign $\pm$ flips in \eqref{eq:RadialGeodesicEquation} and \eqref{eq:PolarGeodesicEquation}, respectively.  The radial potential is quartic in $r$.  The closed-form expression for the roots is not particularly helpful.  On the other hand, the $\theta$ turning points [zeroes of $\Theta$] have a simple expression:
\begin{align}
	\label{eq:AngularTurningPoints}
	\theta_\pm=\arccos\pa{\mp\sqrt{\Delta_\theta+\sqrt{\Delta_\theta^2+\frac{\hat{q}^2}{a^2}}}},\qquad
	\Delta_\theta=\frac{1}{2}\pa{1-\frac{\hat{q}^2+\hat{\lambda}^2}{a^2}}.
\end{align}
For photons that reach infinity, the conserved quantity $E$ is equal to the energy measured by stationary observers at infinity.  The energy measured in the rest frame of the emitting source is
\begin{align}
	E_s=p^{(t)}
	=-p_\mu u^\mu
	=\gamma E\sqrt{\frac{\Xi}{\Delta\Sigma}}\pa{1-\Omega_s\hat{\lambda}}.
\end{align}
The ratio is the ``redshift factor'' $g$, given by
\begin{align}
	\label{eq:Redshift}
	g=\frac{E}{E_s}
	=\frac{1}{\gamma}\sqrt{\frac{\Delta\Sigma}{\Xi}}\frac{1}{1-\Omega_s\hat{\lambda}}
	=\frac{\sqrt{r_s^3-3Mr_s^2\pm2aM^{1/2}r_s^{3/2}}}{r_s^{3/2}\pm M^{1/2}\pa{a-\hat{\lambda}}},
\end{align}
where again the upper/lower sign corresponds to a prograde/retrograde orbit.  Notice that the redshift depends only on $\hat{\lambda}$ and not $\hat{q}$.

In general, the system of equations \eqref{eq:GeodesicEquation} cannot be solved in closed form, and must be approximated numerically.  In Sec.~\ref{sec:NearExtremalExpansion}, we will find tremendous simplifications in the high-spin limit, which will allow us to proceed mostly analytically.  Moreover, we will see in Sec.~\ref{sec:ObservationalAppearance} that these solutions exhibit a variety of surprising observable phenomena not previously encountered in numerical studies.

The conserved quantities $\hat{\lambda}$ and $\hat{q}$ help connect the angle of emission to the angle of reception or equivalently, the image position on the screen.  We parameterize the emission angle by $(\Theta,\Phi)$ defined as the direction cosines in the local rest frame\footnote{That is, we denote by $\Phi$ the angle that the photon three-velocity in the rest frame of the emitter makes with the direction of motion ($\phi$-direction), and we denote by $\Theta$ the angle relative to the local azimuth perpendicular to the equatorial plane ($-\theta$-direction).}
\begin{align}
	\label{eq:DirectionCosines}
	\cos{\Phi}=\frac{p^{(\phi)}}{p^{(t)}}
	=\frac{\pa{\Sigma/\Xi}\sqrt{\Delta}\hat{\lambda}-v_s\pa{1-\omega\hat{\lambda}}}{1-\Omega_s\hat{\lambda}},\qquad
	\cos{\Theta}=-\frac{p^{(\theta)}}{p^{(t)}}
	=\mp\frac{\hat{q}g}{r_s},
\end{align}
where the upper/lower sign corresponds to that in Eq.~\eqref{eq:PolarGeodesicEquation}.  Inverting these relations gives $\hat{\lambda}$ and $\hat{q}$ as a function of the emission angles,
\begin{align}
	\label{eq:ConservedQuantitiesDirectionCosines}
	\hat{\lambda}=\frac{\cos{\Phi}+v_s}{\pa{\Sigma/\Xi}\sqrt{\Delta}+\Omega_s\cos{\Phi}+\omega v_s},\qquad
	\hat{q}=\mp\frac{r_s\cos{\Theta}}{g},
\end{align}
where again the upper/lower sign corresponds to that in Eq.~\eqref{eq:PolarGeodesicEquation} (and ensures $\hat{q}>0$).  Photons with $\theta\to\theta_o$ as $r\to\infty$ correspond to an image of the emitter on the observer's screen.  Here and hereafter, the subscript $o$ stands for ``observer."  The angle of approach to $\theta_o$ corresponds to the position of the image on the observer screen.  Following Refs.~\cite{Bardeen1973}, we use ``screen coordinates'' $\pa{\alpha,\beta}$ corresponding to the apparent position on the plane of the sky.  As we review in App.~\ref{app:ScreenCoordinates}, these are related to the conserved quantities by
\begin{align}
	\label{eq:ScreenCoordinates}
	\alpha=-\frac{\hat{\lambda}}{\sin{\theta_o}},\qquad
	\beta=\pm\sqrt{\hat{q}^2+a^2\cos^2{\theta_o}-\hat{\lambda}^2\cot^2{\theta_o}}
	=\pm\sqrt{\Theta(\theta_o)}.
\end{align}
The sign $\pm$ is equal to the sign of $p_\theta$ (the $\theta$ component of the photon four-momentum) at the observer, which determines whether the photon arrives from above or below.  The angles on the observer sky are given by $\alpha/r_o$ and $\beta/r_o$, where $r_o$ is the distance to the source.

\subsection{Image positions}

Integrating up Eqs.~\eqref{eq:GeodesicEquation} reduces the geodesic equation to quadratures.  That is, the geodesic(s) connecting a source $(t_s,r_s,\theta_s,\phi_s)$ to an observer $(t_o,r_o,\theta_o,\phi_o)$ satisfy\footnote{In the special case of a completely equatorial geodesic, Eq.~\eqref{eq:RTheta} must be discarded.  The trajectory is instead governed by Eqs.~\eqref{eq:DeltaPhi}--\eqref{eq:DeltaT} without the angular integrals.  We do not consider equatorial geodesics in this paper, as they are only relevant for a measure-zero set of observers.}
\begin{subequations}
\label{eq:RaytracingEquation}
\begin{align}
	\label{eq:RTheta}
	&\fint_{r_s}^{r_o}\frac{\ed r}{\pm\sqrt{\mathcal{R}(r)}}
	=\fint_{\theta_s}^{\theta_o}\frac{\ed\theta}{\pm\sqrt{\Theta(\theta)}},\\
	\label{eq:DeltaPhi}
	\Delta\phi=\phi_o-\phi_s=&\fint_{r_s}^{r_o}\frac{a}{\pm\Delta\sqrt{\mathcal{R}(r)}}\pa{2Mr-a\hat{\lambda}}\ed r
	+\fint_{\theta_s}^{\theta_o}\frac{\hat{\lambda}\csc^2{\theta}}{\pm\sqrt{\Theta(\theta)}}\ed\theta,\\
	\label{eq:DeltaT}
	\Delta t=t_o-t_s=&\fint_{r_s}^{r_o}\frac{r}{\pm\Delta\sqrt{\mathcal{R}(r)}}\br{r^3+a^2\pa{r+2M}-2aM\hat{\lambda}}\ed r
	+\fint_{\theta_s}^{\theta_o}\frac{a^2\cos^2{\theta}}{\pm\sqrt{\Theta(\theta)}}\ed\theta.
\end{align}
\end{subequations}
The slash notation $\fint$ indicates that these integrals are to be considered line integrals along a trajectory connecting the two points, where turning points in $r$ or $\theta$ occur any time the corresponding potential $\mathcal{R}(r)$ or $\Theta(\theta)$ vanishes.  The signs $\pm$ in front of $\sqrt{\mathcal{R}(r)}$ and $\sqrt{\Theta(\theta)}$ are chosen to be the same as that of $\ed r$ and $\ed\theta$, respectively.  Each solution of Eqs.~\eqref{eq:RaytracingEquation} corresponds to a null geodesic (labeled by $\hat{\lambda}$ and $\hat{q}$) connecting the source point to the observer point.  For any given pair of points, there may be no solutions, a single solution, or many solutions.

The problem has an equatorial reflection symmetry. Without loss of generality, we take the observer to sit in the northern hemisphere $\theta_o\in\pa{0,\pi/2}$.  We exclude the measure-zero (and mathematically inconvenient) cases of an exactly face-on $(\theta_o=0)$ or edge-on $(\theta_o=\pi/2)$ observation.  We place the observer at angular coordinate $\phi_o=0$ for all time $t_o$, while we place the source at angular coordinate $\phi_s$ at the initial time $t_s=0$.  The coordinates of the source and observer are thus chosen and interpreted as follows:
\begin{subequations}
\label{eq:GeodesicLegs}
\begin{align}
	t_s&:\text{emission time},&
	t_o&:\text{reception time},\\
	r_s&:\text{orbital radius},&
	r_o&\to\infty,\\
	\theta_s&=\pi/2,&
	\theta_o&\in\pa{0,\pi/2}:\text{inclination angle},\\
	\phi_s&=\Omega_st_s,&
	\phi_o&=0.
\end{align}
\end{subequations}
The images seen at inclination angle $\theta_o$ of a source orbiting at $r_s$ may thus be determined as follows:  For each observer time $t_o$, one makes the choices \eqref{eq:GeodesicLegs} for $r_s,\theta_s,\phi_s,r_o,\theta_o,\phi_o$ and plugs them into the basic equations \eqref{eq:RaytracingEquation}.  This produces a set of three integral equations for three variables $t_s,\hat{\lambda},\hat{q}$ in terms of $t_o$.  Each solution then corresponds to an image whose location is fixed by $\hat{\lambda}$ and $\hat{q}$ using Eq.~\eqref{eq:ScreenCoordinates}.  (The emission time $t_s$ may be computed, but is not of observable interest.  We will decouple it from the equations, so that we solve two equations for the two variables $\hat{\lambda},\hat{q}$.)  Solving this problem as a function of the time $t_o$ provides the time-dependent positions of the images.  The total image will be periodic with periodicity equal to that of the source ($T_s=2\pi/\Omega_s$).  However, individual images may evolve on longer timescales (see Fig.~\ref{fig:3x3}), with the required periodicity emerging only after the totality of images is summed over.

We now make the ray-tracing equations \eqref{eq:RaytracingEquation} more explicit by introducing labels to account for the number of rotations ($i.e.$, increases of $\phi$ by $2\pi$) and librations ($i.e.$, turning points in $\theta$).  See Ref.~\cite{Vazquez2004} for a similar treatment.  For the winding in $\phi$, we could introduce an integral winding number $n$, $i.e.$, $n=\mod_{2\pi}\Delta\phi$.  However, we find it more convenient to instead allow the observation point $\phi_o=0$ to take on any physically equivalent value, $i.e.$, $\phi_o=2\pi N$ for an integer $N$.  Using $\phi_s=\Omega_st_s$, we then have
\begin{align}
	2\pi N=\phi_o
	=\Delta\phi+\phi_s
	=\Delta\phi+\Omega_st_s
	=\Delta\phi-\Omega_s\Delta t+\Omega_st_o,
\end{align}
which implies
\begin{align}
	\label{eq:RaytracingCondition}
	\Delta\phi-\Omega_s\Delta t=-\Omega_st_o+2\pi N.
\end{align}
This form is natural because $\phi-\Omega_st$ is conjugate to the $\partial_t+\Omega_s\partial_\phi$ symmetry of the problem.  Note that the periodicity of the image is manifest in that $N\to N+1$ absorbs the shift $t_o\to t_o+T_s$ (with $T_s=2\pi/\Omega_s$), leaving Eq.~\eqref{eq:RaytracingCondition} invariant.  To fix the physical meaning of $N$, we restrict to a single period $t_o\in\pa{0,T_s}$.  Then 
$N$ tracks the number of \textit{extra} windings executed by the photon relative to the emitter between its time of emission and reception.\footnote{The winding number of the photon trajectory is $n=\mod_{2\pi}\Delta\phi$. In the time interval $\br{t_s,t_o}$, the emitter undergoes $n_s=\mod_{2\pi}\Omega_s\Delta t$ windings.  Since $t_o\in\pa{0,T_s}$ by assumption, $\mod_{2\pi}t_o=0$ and therefore Eq.~\eqref{eq:RaytracingCondition} implies that $N=n-n_s$.}  Note that in the near-horizon, near-extremal limit below, $\Delta t$ and $\Delta \phi$ (and hence the winding number $n$) separately diverge linearly [Eq.~\eqref{eq:DeltaPhiLimit}], with the combination $\Delta\phi-\Omega_s\Delta t$ and its net winding number $N$ displaying a milder log divergence.  This reflects the fact that photons received at a fixed time can have been emitted arbitrarily far in the past.

For the radial turning points, we note that there are two possibilities for light reaching infinity: ``direct'' trajectories that are initially outward-bound and have no turning points, and ``reflected'' trajectories that are initially inward-bound but have one radial turning point.  We label these by $b=0$ (direct) and $b=1$ (reflected).  For the $\theta$ turning points, we let $m\geq0$ denote the number of turning points and $s\in\cu{+1,-1}$ denote the \textit{final} sign of $p_\theta$, which is equal to the sign of $\beta$ in Eq.~\eqref{eq:ScreenCoordinates}.

Putting everything together, the basic equations \eqref{eq:RTheta} and \eqref{eq:RaytracingCondition} can be re-expressed as the ``Kerr lens equations''
\begin{subequations}
\label{eq:LensEquations}
\begin{align}
	\label{eq:LensEquation1}
	I_r+b\tilde{I}_r&=G_\theta^{m,s},\\
	\label{eq:LensEquation2}
	J_r+b\tilde{J}_r+\frac{\hat{\lambda}G^{m,s}_\phi-\Omega_sa^2G^{m,s}_t}{M}&=-\Omega_st_o+2\pi N,
\end{align}
\end{subequations}
where the factor of $M$ was introduced to make both equations dimensionless, and we defined
\begin{align}
	G^{m,s}_i=
	\begin{cases}
		\hat{G}_i\qquad\qquad&m=0,\\
		mG_i-s\hat{G}_i\qquad&m\ge1,
	\end{cases}
	\qquad
	i\in\cu{t,\theta,\phi},
\end{align}
with
\begin{subequations}
\begin{align}
	G_\theta&=M\int_{\theta_-}^{\theta_+}\frac{\ed\theta}{\sqrt{\Theta(\theta)}},&
	\hat{G}_\theta&=M\int_{\theta_o}^{\pi/2}\frac{\ed\theta}{\sqrt{\Theta(\theta)}},\\
	G_\phi&=M\int_{\theta_-}^{\theta_+}\frac{\csc^2{\theta}}{\sqrt{\Theta(\theta)}}\ed\theta,&
	\hat{G}_\phi&=M\int_{\theta_o}^{\pi/2}\frac{\csc^2{\theta}}{\sqrt{\Theta(\theta)}}\ed\theta,\\
	G_t&=M\int_{\theta_-}^{\theta_+}\frac{\cos^2{\theta}}{\sqrt{\Theta(\theta)}}\ed\theta,&
	\hat{G}_t&=M\int_{\theta_o}^{\pi/2}\frac{\cos^2{\theta}}{\sqrt{\Theta(\theta)}}\ed\theta,
\end{align}
\end{subequations}
and
\begin{subequations}
\begin{align}
	I_r&=M\int_{r_s}^{r_o}\frac{\ed r}{\sqrt{\mathcal{R}(r)}},&
	\tilde{I}_r&=2M\int_{r_\mathrm{min}}^{r_s}\frac{\ed r}{\sqrt{\mathcal{R}(r)}},\\
	J_r&=\int_{r_s}^{r_o}\frac{\mathcal{J}_r}{\sqrt{\mathcal{R}(r)}}\ed r,&
	\tilde{J}_r&=2\int^{r_s}_{r_\mathrm{min}}\frac{\mathcal{J}_r}{\sqrt{\mathcal{R}(r)}}\ed r,\\
	\label{eq:RadialIntegrand}
	\mathcal{J}_r&=\frac{a\pa{2Mr-a\hat{\lambda}}-\Omega_sr\br{r^3+a^2\pa{r+2M}-2aM\hat{\lambda}}}{\Delta},
\end{align}
\end{subequations}
where $r_\mathrm{min}$ is the largest (real) root of $\mathcal{R}(r)$.  These equations are valid when $r_\mathrm{min}<r_s$, which is always true for light that can reach infinity.  The $G$ integrals are expressed in terms of elliptic functions in Ref.~\cite{InProgress} and reproduced for convenience in App.~\ref{app:AngularIntegrals}.  In general, the $I$ and $J$ integrals can only be computed numerically, but we will compute them analytically in the near-extremal regime considered in Sec.~\ref{sec:NearExtremalExpansion}.

To summarize: for each choice of net winding number $N\in\mathbb{Z}$, polar angular turning points $m\in\mathbb{Z}^{\ge0}$, final vertical orientation $s\in\cu{+1,-1}$, and radial turning point number $b\in\cu{0,1}$, each solution of Eqs.~\eqref{eq:LensEquations} corresponds to a null geodesic connecting source to observer.  The image position $\pa{\alpha,\beta}$ is given by Eq.~\eqref{eq:ScreenCoordinates} with the sign of $\beta$ equal to $s$.

\subsection{Image fluxes}

Each null geodesic connecting the source to the observer produces an apparent image of the source on the observer's screen.  The total flux (energy per unit time) of each image depends on the properties of the source, as well as the lensing effects of gravity.  In order to extract only the effects of gravity, we normalize relative to the comparable ``Newtonian flux'': the flux from the same source at a distance $r_o$ in flat spacetime.

Following Ref.~\cite{Cunningham1973}, we consider a sphere of proper radius $\rho\ll M$ emitting steadily and isotropically with intensity (flux per unit solid angle) $I_s$ in its rest frame.  The Newtonian answer in this case is
\begin{align}
	F_N=I_s\frac{\pi\rho^2}{r_o^2}.
\end{align}
The formalism for computing the relativistic flux was developed in Ref.~\cite{Cunningham1973} in the special case of an extremal black hole ($a=M$).  We now generalize to include the case $a\neq M$.  The intensity of a narrow bundle of light rays varies as the fourth power of the redshift (see, $e.g.$, Ref.~\cite{Johnson1982}),
\begin{align}
	I_o=g^4I_s,
\end{align}
where $g$ is the redshift factor defined in Eq.~\eqref{eq:Redshift}.  The total flux is the integral of the observed intensity over the solid angle subtended by the bundle of rays when it reaches the observer screen.  Noting that $\ed\alpha\ed\beta/r_o^2$ is the area element of solid angle, the flux is thus 
\begin{align}
	F_o=\oiint\frac{\ed\alpha\ed\beta}{r_o^2}g^4 I_s.
\end{align}
In general, $g$ depends on the angle of emission from the emitter.  We consider $\rho\ll1$ and hence may consider rays that deviate only infinitesimally from the central one.  In this approximation, $g$ is constant over the image and, as long as we are restricting to a single image, may be pulled out of the integral, allowing the flux for that image to be written as
\begin{align}
	\frac{F_o}{F_N}=\frac{g^4}{\pi\rho^2}\mathcal{A},\qquad
	\mathcal{A}=\oiint\ed\alpha\ed\beta,
\end{align}
where $\mathcal{A}$ is now the physical area of the apparent image on the plane of the sky, as defined in App.~\ref{app:ScreenCoordinates}.  The coordinates $\pa{\alpha,\beta}$ correspond to the arrival position of each light ray.  To compute the area, we instead change to coordinates $(Y_s,Z_s)$ that characterize the emergence of the light ray from the source.  Equation \eqref{eq:EmitterFrame} provides a locally Minkowski coordinate system $(T,X,Y,Z)$ with origin at the center of the emitter,
\begin{subequations}
\label{eq:Deviation}
\begin{align}
	t-t_\star&=\gamma\sqrt{\frac{\Xi}{\Delta\Sigma}}\pa{T+v_sY},\qquad
	r-r_\star=\sqrt{\frac{\Delta}{\Sigma}}X,\qquad
	\theta-\theta_\star=-\frac{1}{\sqrt{\Sigma}}Z,\\
	\phi-\phi_\star&=\gamma\sqrt{\frac{\Xi}{\Delta\Sigma}}\pa{\Omega_sT+\omega v_sY}+\gamma\sqrt{\frac{\Sigma}{\Xi}}Y,
\end{align}
\end{subequations}
where $x^\mu_\star$ are the coordinates of the point emitter (center of the spherical emitter).

We call the surface $T=X=0$ the ``source screen'' and denote by $(Y_s,Z_s)$ the position of intersection of a light ray with the source screen.\footnote{Note that the source screen is the plane passing through the center of the emitter and lying orthogonal to the Boyer-Lindquist radial direction, at radius $r=r_s$.  In particular, it lies inside of the spherical emitter.  If one prefers to imagine light leaving from the surface of the emitter, these rays must be continued back into the emitter to determine their values of $Y_s,Z_s$.}  In terms of these coordinates, the area of the image takes the form
\begin{align}
	\mathcal{A}=\oiint\ab{\frac{\pd\pa{\alpha,\beta}}{\pd\pa{Y_s,Z_s}}}\ed Y_s\ed Z_s
	=\ab{\frac{\pd\pa{\alpha,\beta}}{\pd\pa{Y_s,Z_s}}}\oiint\ed Y_s\ed Z_s,
\end{align}
where in the second step we note that the Jacobian may be considered constant to leading order, since the region of integration shrinks to zero size as $\rho\to0$.  It remains to compute the Jacobian and the area $\oiint\ed Y_s\ed Z_s$ on the source screen.

To leading order in $\rho$, the bundle of rays forming the image all leave the emitter's surface in the same direction as the central geodesic.  A unit vector in this direction of propagation is given by
\begin{align}
	\hat{k}=\frac{1}{p^{(t)}}\pa{p^{(r)}\hat{X}+p^{(\phi)}\hat{Y}-p^{(\theta)}\hat{Z}}.
\end{align}
Here, $p^{(\phi)}/p^{(t)}$ and $-p^{(\theta)}/p^{(t)}$ are the direction cosines introduced above in Eq.~\eqref{eq:DirectionCosines}, and using Eqs.~\eqref{eq:EmitterFrame} and \eqref{eq:GeodesicEquation} we may compute
\begin{align}
	\frac{p^{(r)}}{p^{(t)}}=\pm g\sqrt{\frac{\mathcal{R}(r_s)}{\Sigma(r_s,\theta_s)\Delta(r_s)}},
\end{align}
where the upper/lower sign corresponds to that in Eq.~\eqref{eq:RadialGeodesicEquation}.  The rays leave the emitter's surface in the hemisphere defined by $\hat{k}$.  If we follow them backwards into the emitter, they intersect the source screen $X=0$ in an ellipse with area $\pi\rho^2/|\hat{k}\cdot\hat{X}|$ (the projection of the hemisphere onto the screen).  Thus we have
\begin{align}
	\oiint\ed Y_s\ed Z_s=\frac{\pi\rho^2}{\ab{\hat{k}\cdot\hat{X}}}
	=\pi\rho^2\ab{\frac{p^{(t)}}{p^{(r)}}}
	=\frac{\pi\rho^2}{g}\sqrt{\frac{\Sigma(r_s,\theta_s)\Delta(r_s)}{\mathcal{R}(r_s)}}.
\end{align}
It remains to compute the Jacobian between $\pa{\alpha,\beta}$ and $(Y_s,Z_s)$ induced by the geodesic equation.  It will be convenient to do so in three stages,
\begin{align}
\label{eq:JacobianDecomposition}
	\ab{\frac{\pd\pa{\alpha,\beta}}{\pd\pa{Y_s,Z_s}}}
	=\ab{\frac{\pd\pa{\tilde{\phi}_s,\theta_s}}{\pd\pa{Y_s,Z_s}}}
	\ab{\frac{\pd\pa{\tilde{\phi}_s,\theta_s}}{\pd\pa{\hat{\lambda},\hat{q}}}}^{-1}
	\ab{\frac{\pd\pa{\alpha,\beta}}{\pd\pa{\hat{\lambda},\hat{q}}}}.
\end{align}
Here $\tilde{\phi}=\phi-\Omega_s t$, and $\pa{\theta_s,\tilde{\phi}_s}$ refer to the coordinate values at the intersection of the geodesic with the source screen $T=X=0$.  From Eq.~\eqref{eq:Deviation} we have
 \begin{align}
	\theta_s-\theta_\star=-\frac{1}{\sqrt{\Sigma}}Z,\qquad
	\tilde{\phi}_s-\tilde{\phi}_\star=\gamma\pa{\sqrt{\frac{\Xi}{\Delta\Sigma}}\pa{\omega-\Omega_s}v_s+\sqrt{\frac{\Sigma}{\Xi}}}Y,
\end{align}
which allows the first determinant in \eqref{eq:JacobianDecomposition} to be computed as
\begin{align}
	\ab{\frac{\pd\pa{\tilde{\phi}_s,\theta_s}}{\pd\pa{Y_s,Z_s}}}
	=\frac{1}{r_s\pa{r_s^{3/2}\pm aM^{1/2}}}\sqrt{\frac{r_s^3-3Mr_s^2\pm2aM^{1/2}r_s^{3/2}}{\Delta}},
\end{align}
where the upper/lower sign once again corresponds to a prograde/retrograde orbit.  The third determinant in \eqref{eq:JacobianDecomposition} may be computed from Eq.~\eqref{eq:ScreenCoordinates} as
\begin{align}
	\ab{\frac{\pd\pa{\alpha,\beta}}{\pd\pa{\hat{\lambda},\hat{q}}}}
	=\frac{\hat{q}}{\sin{\theta_o}\sqrt{\hat{q}^2+a^2\cos^2{\theta_o}-\hat{\lambda}^2\cot^2{\theta_o}}}
	=\frac{\hat{q}}{\sin{\theta_o}\sqrt{\Theta(\theta_o)}}.
\end{align}
The middle determinant in \eqref{eq:JacobianDecomposition} involves using the geodesic equation to study the variation of $\theta_s$ and $\tilde{\phi}_s$ with $\hat{\lambda}$ and $\hat{q}$.  Recall that $\theta_s$ and $\tilde{\phi}_s$ are the intersection of the geodesic with the source screen $T=X=0$.  We must therefore vary the geodesic equation at fixed source radius $r_s$ (to stay on the screen $X=0$)\footnote{The source time $t_s$ also varies in the manner prescribed by \eqref{eq:Deviation} with $X=T=0$.  However, we will not need the explicit form because we have cast the problem entirely in terms of the special combination $\tilde{\phi}=\phi-\Omega_st$, which is possible because of the co-rotation symmetry.} and at fixed observer position $t_o,r_o,\theta_o,\phi_o$.  We must first generalize Eq.~\eqref{eq:LensEquation1} to allow the source to be outside the equatorial plane, $i.e.$, $\theta_s\neq\pi/2$.  This gives rise to an extra integral from $\theta_s$ to $\pi/2$, and we write the equation
\begin{align}
	\label{eq:A}
	A=0,\qquad
	A\equiv I_r+b\tilde{I}_r-G_\theta^{m,s}\pm M\int_{\pi/2}^{\theta_s}\frac{\ed\theta}{\sqrt{\Theta(\theta)}},
\end{align}
where the upper/lower sign corresponds to pushing the source above/below the equatorial plane, and drops out of the final answer for the flux.  A change in $\hat{\lambda}$ at fixed $\hat{q}$ induces changes in $\theta_s$ and $\tilde{\phi}_s$, and we denote by $d/d\hat{\lambda}$ the total derivative including these changes.  Similarly, we denote by $d/d \hat{q}$ the total derivative at fixed $\hat{\lambda}$.  It is these total derivatives of $A$ which must vanish, 
\begin{subequations}
\label{eq:DerivativesOfA}
\begin{align}
	\left.\frac{dA}{d\hat{\lambda}}\right|_{\hat{q}}&=\frac{\pd A}{\pd\hat{\lambda}}+\frac{\pd\theta_s}{\pd\hat{\lambda}}\frac{\pd A}{\pd\theta_s}+\frac{\pd\tilde{\phi}_s}{\pd\hat{\lambda}}\frac{\pd A}{\pd\tilde{\phi}_s}=0,\\
	\left.\frac{dA}{d\hat{q}}\right|_{\hat{\lambda}}&=\frac{\pd A}{\pd\hat{q}}+\frac{\pd\theta_s}{\pd\hat{q}}\frac{\pd A}{\pd\theta_s}+\frac{\pd\tilde{\phi}_s}{\pd\hat{q}}\frac{\pd A}{\pd\tilde{\phi}_s}=0.
\end{align}
\end{subequations}
From the definition of $A$, we see that
\begin{align}
	\frac{\pd A}{\pd\tilde{\phi}_s}=0,\qquad
	\frac{\pd A}{\pd\theta_s}=\pm\frac{M}{\sqrt{\Theta(\theta_s)}}.
\end{align}
From Eqs.~\eqref{eq:DerivativesOfA}, we then see that
\begin{align}
	\frac{\pd\theta_s}{\pd\hat{\lambda}}=\mp\frac{\frac{\pd A}{\pd\hat{\lambda}}}{{\frac{M}{\sqrt{\Theta(\theta_s)}}}},\qquad
	\frac{\pd\theta_s}{\pd\hat{q}}=\mp\frac{\frac{\pd A}{\pd\hat{q}}}{{\frac{M}{\sqrt{\Theta(\theta_s)}}}}.
\end{align}
The partial derivatives of $\tilde{\phi}_s$ are easier to obtain since [noting Eq.~\eqref{eq:RaytracingCondition}] Eq.~\eqref{eq:LensEquation2} is just a formula for $\Delta \tilde{\phi}$,
\begin{align}
	\label{eq:B}
	\tilde{\phi}_o-\tilde{\phi}_s=B
	=-\Omega_st_o+2\pi N,\qquad
	B\equiv J_r+b\tilde{J}_r+\frac{\hat{\lambda}G^{m,s}_\phi-\Omega_sa^2G^{m,s}_t}{M}.
\end{align}
Taking the partial derivative with respect to $\hat{\lambda}$ and $\hat{q}$ directly gives 
\begin{align}
	\frac{\pd\tilde{\phi}_s}{\pd\hat{\lambda}}=-\frac{\pd B}{\pd\hat{\lambda}},\qquad
	\frac{\pd\tilde{\phi}_s}{\pd\hat{q}}=-\frac{\pd B}{\pd\hat{q}}.
\end{align}
Thus, we at last obtain the middle determinant in Eq.~\eqref{eq:JacobianDecomposition},
\begin{align}
	\ab{\frac{\pd\pa{\tilde{\phi}_s,\theta_s}}{\pd\pa{\hat{\lambda},\hat{q}}}}
	=\frac{\sqrt{\Theta(\theta_s)}}{M}\ab{\det
	\begin{pmatrix}
		\frac{\pd B}{\pd\hat{\lambda}} & \frac{\pd B}{\pd\hat{q}}\vspace{2pt}\\
		\frac{\pd A}{\pd\hat{\lambda}} & \frac{\pd A}{\pd\hat{q}}
	\end{pmatrix}}.
\end{align}
Putting everything together, the flux is given by
\begin{align}
	\label{eq:Flux}
	\frac{F_o}{F_N}=\frac{g^3}{r_s^{3/2}\pm aM^{1/2}}\sqrt{\frac{r_s^3-3Mr_s^2\pm2aM^{1/2}r_s^{3/2}}{\mathcal{R}(r_s)\Theta(\theta_o)}}\frac{M}{\sin{\theta_o}}\ab{\det
	\begin{pmatrix}
		\frac{\pd B}{\pd\hat{\lambda}} & \frac{\pd B}{\pd\hat{q}}\vspace{2pt}\\
		\frac{\pd A}{\pd\hat{\lambda}} & \frac{\pd A}{\pd\hat{q}}
	\end{pmatrix}}^{-1},
\end{align}
where again the upper/lower sign corresponds to a prograde/retrograde orbit, and $A$ and $B$ are given in \eqref{eq:A} and \eqref{eq:B}.

To summarize: each choice of $\hat{\lambda}$ and $\hat{q}$ that solves Eqs.~\eqref{eq:LensEquations} corresponds to an image of the source appearing at time $t_o$ at position given by \eqref{eq:ScreenCoordinates} and flux given by \eqref{eq:Flux}.  The determinant is analytically intractable in the general case, but in the next section, we will compute it explicitly in the near-extremal regime.

\section{Near-extremal expansion}
\label{sec:NearExtremalExpansion}

We now specialize to the case of an emitter orbiting on, or near, the (prograde) Innermost Stable Circular Orbit (ISCO) of a near-extremal ($i.e.$, high-spin)  black hole.  We will work with a dimensionless radial coordinate $R$, simply related to the Boyer-Lindquist radius $r$ by
\begin{align}
	\label{eq:R}
	R=\frac{r-M}{M}.
\end{align}
We introduce a small parameter $\epsilon$ to represent the deviation of the black hole from extremality,
\begin{align}
	\label{eq:aExpansion}
	a=M\sqrt{1-\epsilon^3},\qquad
	\epsilon\ll1.
\end{align}
The choice of the third power of $\epsilon$ in this expression puts the ISCO a coordinate distance $\sim\epsilon$ from the horizon,
\begin{align}
	R_\mathrm{ISCO}=2^{1/3}\epsilon+\O{\epsilon^2}.
\end{align}
Note that the proper distance $d$ from the ISCO to the edge of the throat (defined here as the equatorial edge of the ergosphere, $r=2M$ for all spin) is related to $\epsilon$ at leading order by
\begin{align}
	\epsilon=\frac{e}{2^{1/3}}e^{-d/M}.
\end{align}
This implies that corrections are exponentially suppressed in the distance $d$, which is the expected behavior near a critical point.  That is, the proper size $d$ of the throat acts as the diverging correlation length of the system.

The observer is at coordinate position $R_o=\pa{r_o-M}/M\approx r_o/M$, while the source orbits on or near the ISCO,
\begin{align}
	\label{eq:rExpansion}
	R_s=\epsilon\bar{R}+\O{\epsilon^2},\qquad
	\bar{R}\ge2^{1/3}.
\end{align}
Photons departing from this source have rather constrained values of the conserved quantities $\hat{\lambda}$ and $\hat{q}$.  Plugging into Eq.~\eqref{eq:ConservedQuantitiesDirectionCosines} shows that
\begin{align}
	\hat{\lambda}=2M+\frac{3M\bar{R}\cos{\Phi}}{1+2\cos{\Phi}}\epsilon+\O{\epsilon^2}.
\end{align}
Thus, apart from the measure-zero cases $\cos{\Phi}=-1/2$ and $\cos{\Phi}=0$, $\hat{\lambda}$ approaches $2M$ with corrections scaling like $\epsilon$.  This means that emissions are constrained to be near the superradiant bound.  We can keep track of the small corrections by working with a new quantity $\lambda$ instead of $\hat{\lambda}$,
\begin{align}
	\label{eq:LambdaExpansion}
	\hat{\lambda}=2M\pa{1-\epsilon\lambda}.
\end{align}
Following Ref.~\cite{Porfyriadis2016}, we will also introduce a new quantity $q$ by
\begin{align}
	\label{eq:qExpansion}
	\hat{q}=M\sqrt{3-q^2}.
\end{align}
The screen coordinates are then given by 
\begin{align}
	\label{eq:AlphaBetaExpansion}
	\alpha=-\frac{2M}{\sin{\theta_o}}+\O{\epsilon},\qquad
	\beta=sM\sqrt{3-q^2+\cos^2{\theta_o}-4\cot^2{\theta_o}}+\O{\epsilon}.
\end{align}
Notice that $\lambda$ does not appear to leading order.  The requirement that $\beta$ is real (so that photons can reach infinity) establishes a range of $q$:
\begin{align}
	\label{eq:qRange}
	q\in\br{0,\sqrt{3+\cos^2{\theta_o}-4\cot^2{\theta_o}}}.
\end{align}
Equations~\eqref{eq:AlphaBetaExpansion} and \eqref{eq:qRange} correspond to a vertical line segment we call the NHEKline (App.~\ref{app:Shadow}).  Thus we see that, as $\epsilon \rightarrow 0$, all light ends up on the NHEKline.  Note that there is no range of $q$ at all when $\theta<\theta_\mathrm{crit}\approx47^\circ$, corresponding to the disappearance of the NHEKline (App.~\ref{app:Shadow}).  Interestingly, the geometry of the submanifold $\theta=\theta_\mathrm{crit}$ is precisely 3-dimensional Anti-de Sitter space with radius $\ell=4M\pa{\sqrt{3}-1}^{-1/2}$.

In App.~\ref{app:Shadow}, we show that \textit{all} light from near-horizon sources, not just the equatorial near-ISCO emission considered here, ends up on the NHEKline.  This generalizes and makes precise an old observation of Bardeen \cite{Bardeen1973}.

For later reference, we now expand various formulae in $\epsilon$.  The redshift [Eq.~\eqref{eq:Redshift}] and direction cosines [Eqs.~\eqref{eq:DirectionCosines}] are given by
\begin{align}
	\label{eq:RedshiftLimit}
	g&=\frac{1}{\sqrt{3}+\frac{4}{\sqrt{3}}\frac{\lambda}{\bar{R}}}+\O{\epsilon},\\
	\label{eq:PhiLimit}
	\cos{\Phi}&=-\frac{1}{2+\frac{3}{2}\frac{\bar{R}}{\lambda}}+\O{\epsilon}
	=\frac{\sqrt{3}}{2}\pa{g-\frac{1}{\sqrt{3}}}+\O{\epsilon},\\
	\cos{\Theta}&=\frac{(-1)^{m+1}s}{1+\frac{4}{3}\frac{\lambda}{\bar{R}}}\sqrt{1-\frac{q^2}{3}}+\O{\epsilon} 
	=(-1)^{m+1}sg\sqrt{3-q^2}+\O{\epsilon}.
\end{align}
The factor $(-1)^{m+1}$ arises in relating the final sign of $p^\theta$ ($i.e.$, $s$) to its initial sign [the $\pm$ in Eq.~\eqref{eq:PolarGeodesicEquation}].  Note by Eq.~\eqref{eq:PhiLimit} that $g$ is bounded above,
\begin{align}
	g\le\sqrt{3},
\end{align}
with the largest values of $g$ achieved by photons emitted in a narrow cone around the forward direction ($\cos{\Phi}\sim1$).  The orbital frequency and period are
\begin{align}
	\Omega_s=\frac{1}{2M}+\O{\epsilon},\qquad
	T_s=\frac{2\pi}{\Omega_s}
	=4\pi M+\O{\epsilon}.
\end{align}

\subsubsection*{Precisely extremal problem}

Our goal is to determine image positions and fluxes for a nearly extremal black hole (small $\epsilon$).  This involves plugging Eqs.~\eqref{eq:aExpansion}, \eqref{eq:rExpansion}, \eqref{eq:LambdaExpansion}, and \eqref{eq:qExpansion} into Eq.~\eqref{eq:LensEquations}, and expanding to leading order in $\epsilon$ (keeping $\bar{R}$, $\lambda$, and $q$ fixed).  We note, however, that the results are identical if we just set $a=M$ rather than use Eq.~\eqref{eq:aExpansion}.  Similarly, the formula below for the flux of each image [Eq.~\eqref{eq:FluxExpansion}] is identical (to leading order) in the two cases.  To be completely explicit, the problem is therefore equivalent to considering, to leading order in $\epsilon$,
\begin{subequations}
\label{eq:ExtremalExpansion}
\begin{align}
	a&=M,\\
	r_s&=M\pa{1+\epsilon\bar{R}},\\
	\hat{\lambda}&=2M\pa{1-\epsilon\lambda},\\
	\hat{q}&=M\sqrt{3-q^2}.
\end{align}
\end{subequations}
This set of equations corresponds to an emitter orbiting near the event horizon of a precisely extremal black hole, with $\epsilon\ll1$ playing the role of a scaling parameter.  Thus we learn that the image of an emitter orbiting near the ISCO of a near-extremal black hole is mapped in a simple way to the image of an emitter orbiting near the horizon of a precisely extremal black hole.  The same agreement was noticed for the gravitational waves emitted by such an emitter \cite{Gralla2015}.  The agreement is likely a consequence of the infinite-dimensional conformal symmetry that maps extremal to near-extremal black holes \cite{Guica2009,Porfyriadis2014a}.  Such maps were previously used to compute gravitational waves from plunging particles \cite{Porfyriadis2014a,Porfyriadis2014b,Hadar2015,Hadar2017}.

\subsection{Image positions}

The image positions are determined by solving Eqs.~\eqref{eq:LensEquations}.  As $\epsilon\to0$, both sides of these equations grow as $\log{\epsilon}$, and the leading term determines the growth of $m$.  In order to determine $\lambda$, and hence the leading flux and redshift of each image, we must keep the subleading $\O{\epsilon^0}$ terms in Eqs.~\eqref{eq:LensEquations}.  These terms are also necessary for any quantitative validity at reasonable values of $\epsilon$, for which $\log{\epsilon}$ will be numerically of order a few.

\subsubsection{First equation}

The first equation to solve is Eq.~\eqref{eq:LensEquation1},
\begin{align}
	\label{eq:FirstEquation}
	I_r+b\tilde{I}_r=mG_{\theta}-s\hat{G}_\theta.
\end{align}
The $G$ integrals are expressed as elliptic functions in Ref.~\cite{InProgress} and reproduced in App.~\ref{app:AngularIntegrals}.  The $I$ integrals may be computed under the approximations \eqref{eq:ExtremalExpansion} using the method of matched asymptotic expansions (App.~\ref{app:MAE}).  The results are
\begin{align}
	\label{eq:Ir}
	I_r&=-\frac{1}{q}\log{\epsilon}+\frac{1}{q}\log\br{\frac{4q^4R_o}{\pa{q^2+qD_o+2R_o}\pa{q^2\bar{R}+qD_s+4\lambda}}}+\O{\epsilon},\\
	\tilde{I}_r&=\frac{1}{q}\log\br{\frac{\pa{q^2\bar{R}+qD_s+4\lambda}^2}{4\pa{4-q^2}\lambda^2}}+\O{\epsilon},
\end{align}
where we introduced
\begin{align}
	\label{eq:RadialIntegrandLimits}
	D_s=\sqrt{q^2\bar{R}^2+8\lambda\bar{R}+4\lambda^2},\qquad
	D_o=\sqrt{q^2+4R_o+R_o^2}.
\end{align}
Note that $D_s=0$ corresponds to emission from a radial turning point (see App.~\ref{app:ReflectedRadialIntegral}), for which $\tilde{I}_r$ and $\tilde{J}_r$ vanish.  The LHS of \eqref{eq:FirstEquation} grows logarithmically in $\epsilon$.  This can only be compensated on the RHS by taking $m$ to scale similarly, and it is convenient to define
\begin{align}
	\label{eq:m}
	m=-\frac{1}{qG_\theta}\log{\epsilon}+\bar{m}.
\end{align}
The first term ensures that \eqref{eq:FirstEquation} is satisfied at leading order $\O{\log{\epsilon}}$, and $\bar{m}$ can be viewed as the $\O{\epsilon^0}$ correction.\footnote{\label{fn:ContinuousLabels} If $m$ is considered a continuous parameter, then we may take $\bar{m}$ to be truly constant (independent of $\epsilon$).  However, in reality $m$ must take integral values, so $\bar{m}$ must vary with $\epsilon$.  We can write $\bar{m}=\bar{m}_0-\delta\bar{m}(\epsilon)$, where $\bar{m}_0$ is an integer and $\delta\bar{m}(\epsilon)=\mod_1\br{-1/(qG_\theta)\log{\epsilon}}$.  Then each choice of integer $\bar{m}_0$ defines $m(\epsilon)$ taking only properly integral values.  We may still regard $\bar{m}(\epsilon)$ as $\O{\epsilon^0}$ since its variation is bounded by unity.}  Plugging \eqref{eq:m} into \eqref{eq:FirstEquation} yields the subleading part of the equation,
\begin{align}
	\frac{1}{q}\log\br{\frac{4q^4R_o}{\pa{q^2+qD_o+2R_o}\pa{q^2\bar{R}+qD_s+4\lambda}}}+\frac{b}{q}\log\br{\frac{\pa{q^2\bar{R}+qD_s+4\lambda}^2}{4\pa{4-q^2}\lambda^2}}=G_\theta^{\bar{m},s}.
\end{align}
Exponentiating this equation yields
\begin{align}
	\label{eq:SimplifiedFirstEquation}
	\frac{4\Upsilon}{q^2\bar{R}+qD_s+4\lambda}\br{\frac{\pa{q^2\bar{R}+qD_s+4\lambda}^2}{4\pa{4-q^2}\lambda^2}}^b=1,
\end{align}
where we introduced $\Upsilon>0$ defined by
\begin{align}
	\label{eq:Upsilon}
	\Upsilon\equiv\frac{q^4R_o}{q^2+qD_o+2R_o}e^{-qG_\theta^{\bar{m},s}}=\frac{q^4}{q+2}e^{-qG_\theta^{\bar{m},s}}+\O{\frac{1}{R_o}}.
\end{align}

We now consider the direct $(b=0)$ and reflected $(b=1)$ cases separately.  For direct images ($b=0$), we can rearrange \eqref{eq:SimplifiedFirstEquation} to give
\begin{align}
	\label{eq:DirectFirstEquation}
	q^2\bar{R}+4\pa{\lambda-\Upsilon}=-qD_s.
\end{align}
Squaring both sides produces a quadratic equation in $\lambda$,
\begin{align}
	\label{eq:DirectFirstEquationSquared}
	\pa{4-q^2}\lambda^2-8\Upsilon\lambda-2\Upsilon\pa{q^2\bar{R}-2\Upsilon}=0,
\end{align}
whose solutions are
\begin{align}
	\label{eq:LambdaSolutions}
	\lambda_\pm&=\frac{2\Upsilon}{4-q^2}\br{2\pm q\sqrt{1+\frac{\bar{R}}{2\Upsilon}\pa{4-q^2}}}.
\end{align}
These solutions to the squared equation \eqref{eq:DirectFirstEquationSquared} will only solve the original equation \eqref{eq:DirectFirstEquation} when its LHS is negative.\footnote{In the special case that $D_s=0$, corresponding to an emitter orbiting precisely at the photon's radial turning point, the equation is solved when the LHS is zero.  We will consider this measure-zero case in Sec.~\ref{sec:m}, but for the time being we can exclude it and thereby make Eq.~\eqref{eq:AdditionalCondition} a strict inequality.}  Thus, an additional condition on a valid solution $\lambda_\pm$ is
\begin{align}
	\label{eq:AdditionalCondition}
	\lambda_\pm<\Upsilon-\frac{q^2\bar{R}}{4}.
\end{align}
However, Eq.~\eqref{eq:LambdaSolutions} shows that $\lambda_+$ is always larger than $\Upsilon>0$, and hence can never satisfy \eqref{eq:AdditionalCondition}.  Thus, only $\lambda_-$ can be a solution.  Plugging the formula \eqref{eq:LambdaSolutions} for $\lambda_-$ into \eqref{eq:AdditionalCondition}, we find that the condition for $\lambda_-$ to be valid is
\begin{align}
	\bar{R}\in\br{z_-,z_+},\qquad
	z_\pm=\frac{4\Upsilon}{q^2}\pa{1\pm\frac{2}{\sqrt{4-q^2}}}\gtrless0.
\end{align}
However, recall that $\bar{R}>0$ in the extremal case and $\bar{R}\ge2^{1/3}$ in the near-extremal case.  Thus we can simplify the condition to
\begin{align}
	\label{eq:DirectCondition}
	\bar{R}<\frac{4\Upsilon}{q^2}\pa{1+\frac{2}{\sqrt{4-q^2}}}.
\end{align}

For reflected images with $b=1$, we can rearrange \eqref{eq:SimplifiedFirstEquation} to produce the analog of \eqref{eq:DirectFirstEquation},
\begin{align}
	\label{eq:ReflectedFirstEquation}
	q^2\bar{R}+4\lambda-\frac{\pa{4-q^2}\lambda^2}{\Upsilon}=-qD_s.
\end{align}
Squaring both sides gives a quartic equation,
\begin{align}
	\label{eq:QuarticEquation}
	\frac{\pa{4-q^2}\lambda^2}{\Upsilon^2}\br{\pa{4-q^2}\lambda^2-8\Upsilon\lambda-2\Upsilon\pa{q^2\bar{R}-2\Upsilon}}=0.
\end{align}
The factor in brackets is actually the same quadratic equation as in the direct case \eqref{eq:DirectFirstEquationSquared}.  Thus the solutions to \eqref{eq:QuarticEquation} are $\lambda=0$ and $\lambda_\pm$ defined in \eqref{eq:LambdaSolutions}.  To be a true solution of the original equation \eqref{eq:ReflectedFirstEquation}, the LHS of \eqref{eq:ReflectedFirstEquation} must be strictly negative.  Thus $\lambda=0$ is inadmissible and $\lambda_\pm$ must satisfy
\begin{align}
	\lambda_\pm<\frac{2\Upsilon}{4-q^2}\pa{1-\sqrt{1+\frac{q^2\bar{R}}{4\Upsilon}\pa{4-q^2}}}<0
	\quad\text{or}\quad
	\lambda_\pm>\frac{2\Upsilon}{4-q^2}\pa{1+\sqrt{1+\frac{q^2\bar{R}}{4\Upsilon}\pa{4-q^2}}}>0.
\end{align}
However, when $\lambda>0$, the outermost turning point is inside the horizon [see Eq.~\eqref{eq:TurningRadius}], meaning there can be no reflected image in this case.   [The lack of a valid trajectory for $b=1$ and $\lambda>0$ can also be seen from the failure of $\tilde{J}_r$ to exist---see the $1/\Delta$ factor in Eq.~\eqref{eq:RadialIntegrand}.]  Thus only $\lambda_-$ is admissible and only the first condition can possibly be satisfied.  Plugging in the formula \eqref{eq:LambdaSolutions} for $\lambda_-$, this condition becomes
\begin{align}
	\bar{R}>\frac{4\Upsilon}{q^2}\pa{1+\frac{2}{\sqrt{4-q^2}}},
\end{align}
which is the opposite inequality of the direct condition \eqref{eq:DirectCondition}.  Both of these inequalities are saturated for the boundary case of emission precisely from a photon's radial turning point.

To summarize, for each choice of $m$, $b$, $s$, and $q$, there is either zero or one solution for $\lambda$.  The solution exists provided
\begin{subequations}
\label{eq:LambdaCondition}
\begin{align}
	\bar{R}<\frac{4\Upsilon}{q^2}\pa{1+\frac{2}{\sqrt{4-q^2}}}
	&\qquad\text{if $b=0$ (direct)},\\
	\bar{R}>\frac{4\Upsilon}{q^2}\pa{1+\frac{2}{\sqrt{4-q^2}}}
	&\qquad\textrm{if $b=1$ (reflected)},
\end{align}
\end{subequations}
in which case it is given by [repeating Eq.~\eqref{eq:LambdaSolutions}, choosing the minus branch]
\begin{align}
	\label{eq:Lambda}
	\lambda=\frac{2\Upsilon}{4-q^2}\br{2-q\sqrt{1+\frac{\bar{R}}{2\Upsilon}\pa{4-q^2}}},\qquad
	\Upsilon=\frac{q^4}{q+2}e^{-qG_\theta^{\bar{m},s}}. 
\end{align}
We have included the large-$R_o$ expression for $\Upsilon$, making the expressions independent of $R_o$.  The full version of $\Upsilon$ is given above in Eq.~\eqref{eq:Upsilon}.  Note that one may equivalently choose $m$, $s$, and $q$, and hence determine $b$ from Eq.~\eqref{eq:LambdaCondition}.

\subsubsection{Second equation}

The second equation to solve is Eq.~\eqref{eq:LensEquation2}.  We will work with a dimensionless time coordinate $\hat{t}_o$ in terms of which the emitter has unit periodicity,
\begin{align}
	\hat{t}_o=\frac{t_o}{T_s}=\frac{t_o}{4\pi M}+\O{\epsilon}.
\end{align}
In terms of this phase, Eq.~\eqref{eq:LensEquation2} can be rewritten as
\begin{align}
	\label{eq:SecondEquation}
	\hat{t}_o=N+\mathcal{G},\qquad
	\mathcal{G}\equiv-\frac{1}{2\pi}\pa{J_r+b\tilde{J}_r+2G^{m,s}_\phi-\frac{1}{2}G^{m,s}_t}.
\end{align}
Without loss of generality, we restrict to the single period $\hat{t}_o\in\br{0,1}$.

The $G$ integrals are given as elliptic functions in Ref.~\cite{InProgress} and reproduced in App.~\ref{app:AngularIntegrals}.  The $J$ integrals may be computed using the method of matched asymptotic expansions (App.~\ref{app:MAE}), giving
\begin{align}
	\label{eq:Jr}
	J_r&=-\frac{7}{2}I_r+\frac{q}{2}\pa{1-\frac{3}{4}\frac{\bar{R}}{\lambda}}-\frac{1}{2}\pa{D_o-\frac{3}{4}\frac{D_s}{\lambda}}+\log\br{\frac{\pa{q+2}^2\bar{R}}{\pa{D_o+R_o+2}\pa{D_s+2\bar{R}+2\lambda}}}+\O{\epsilon},\\
	\label{eq:ReflectedJr}
	\tilde{J}_r&=-\frac{7}{2}\tilde{I}_r-\frac{3}{4}\frac{D_s}{\lambda}+\log\br{\frac{\pa{D_s+2\bar{R}+2\lambda}^2}{\pa{4-q^2}\bar{R}^2}}+\O{\epsilon}.
\end{align}
Note that this expression for $\tilde{J}_r$ is only valid for $\lambda<0$.  When $\lambda>0$, the radial turning point $r_\mathrm{min}$ is inside the horizon (see App.~\ref{app:ReflectedRadialIntegral}) and the integral for $\tilde{J}_r$ does not exist [see the $1/\Delta$ factor in Eq.~\eqref{eq:RadialIntegrand}].  This corresponds to the fact that all reflected light $(b=1)$ has $\lambda<0$, which shows that $\lambda>0$ lies within the shadow.

For each choice of discrete parameters $m,s,b$ having a non-zero range of $q$ satisfying the condition \eqref{eq:LambdaCondition}, $\lambda(q)$ is determined by \eqref{eq:Lambda} and $\mathcal{G}$ becomes a function of $q$.  Equation~\eqref{eq:SecondEquation} then gives the observation time $t_o(q)$ for each choice of an integer $N$.  Restricting to $0\leq\hat{t}_o<1$ determines $N$ uniquely for each $q$, and the multivalued inverse $q\pa{\hat{t}_o}$ corresponds to the tracks of images moving along the NHEKline.  That is, if there are $p$ domains $\br{\bar{t}^{\rm min}_p,\bar{t}_p^{\rm max}}$ where $\hat{t}_o(q)$ is invertible and lies between $0$ and $1$, then each corresponding inverse $q_p\pa{\hat{t}_o}$, taken over its corresponding range $0<\bar{t}^\mathrm{min}_p<\hat{t}_o<\bar{t}_p^\mathrm{max}<1$, describes the track of an image over one period (or just a portion thereof).  The number of inverses changes at local maxima and minima of $\mathcal{G}(q)$, corresponding to a change in the number of images at the associated time $\hat{t}_o$.  Minima correspond to pair creation of images, while maxima correspond to pair annihilation.  Finding the tracks of all such images for all choices of $N,m,s,b$ completes the task of finding the time-dependent locations of the images.  We describe a practical approach, along with an example, in Sec.~\ref{sec:ObservationalAppearance} below.

\subsubsection{Winding number around the axis of symmetry}

The winding number around the axis of symmetry for a photon trajectory is $n=\mod_{2\pi}\Delta\phi$, where $\Delta\phi$ is computed from Eq.~\eqref{eq:DeltaPhi}.  Using the method of matched asymptotic expansions described in App.~\ref{app:MAE}, we find that to leading order in $\epsilon$,
\begin{align}
	\label{eq:DeltaPhiLimit}
	\Delta\phi=\frac{1}{2\lambda\epsilon}\pa{\frac{D_s}{\bar{R}}-q}+\O{\log{\epsilon}}.
\end{align}
Note that this leading order expression diverges linearly in $\epsilon$.

\subsubsection{Scaling with $\epsilon$ and $R_o$}

It is instructive to examine the scaling of various quantities as $\epsilon\to0$ and $R_o\to\infty$.  Plugging in \eqref{eq:m} for $m$ into the definition \eqref{eq:SecondEquation} of $\mathcal{G}$, we find 
\begin{align}
	\label{eq:G}
	\mathcal{G}=-\frac{1}{2\pi}\br{J_r+b\tilde{J}_r-\frac{1}{qG_\theta}\pa{2G_\phi-\frac{1}{2}G_t}\log{\epsilon}+2G^{\bar{m},s}_\phi-\frac{1}{2}G^{\bar{m},s}_t}.
\end{align}
The integral $\tilde{J}_r$ [Eq.~\eqref{eq:ReflectedJr}] is finite for all parameter values, while the integral $J_r$ [Eq.~\eqref{eq:Jr}] diverges as $\epsilon\to0$ and as $R_o\to\infty$,
\begin{align}
	J_r&=\frac{q}{2}\pa{1-\frac{3}{4}\frac{\bar{R}}{\lambda}}-\pa{1-\frac{3}{8}\frac{D_s}{\lambda}}-\frac{7}{2q}\log\br{\frac{4q^4}{\pa{q+2}\pa{q^2\bar{R}+qD_s+4\lambda}}}+\log\br{\frac{\pa{q+2}^2\bar{R}}{2\pa{D_s+2\bar{R}+2\lambda}}}\nonumber\\
	&\quad+\frac{7}{2q}\log{\epsilon}-\pa{\frac{R_o}{2}+\log{R_o}}+\O{\frac{1}{R_o}}+\O{\epsilon}.
\end{align}
Thus Eq.~\eqref{eq:G} has logarithmic divergences in $\epsilon\to0$ as well as linear and logarithmic divergences in $R_o\to\infty$.  These signal that the integer $N$ will become asymptotically large.  Similarly to Eq.~\eqref{eq:m} for $m$, we may define
\begin{align}
	\label{eq:N}
	N=\frac{1}{2\pi}\br{\frac{1}{2q}\pa{7-\frac{4G_\phi-G_t}{G_\theta}}\log{\epsilon}-\pa{\frac{R_o}{2}+\log{R_o}}}+\bar{N},
\end{align}
where $\bar{N}$ can be regarded as $\O{\epsilon^0}$ and $\O{R_o^0}$ (see footnote \ref{fn:ContinuousLabels}).  Plugging \eqref{eq:N} into \eqref{eq:G} gives an equation with all terms $\O{\epsilon^0}$,
\begin{align}
	\hat{t}_o=\bar{N}-\frac{1}{2\pi}\Bigg\{&
	\frac{q}{2}\pa{1-\frac{3}{4}\frac{\bar{R}}{\lambda}}-\pa{1+\pa{2b-1}\frac{3}{8}\frac{D_s}{\lambda}}-\frac{7}{2q}\log\br{\frac{4q^4}{\pa{q+2}}\frac{\pa{q^2\bar{R}+qD_s+4\lambda}^{2b-1}}{\br{4\pa{4-q^2}\lambda^2}^b}}\nonumber\\
	&\quad+\log\br{\frac{\pa{q+2}^2}{2\pa{4-q^2}^b}\pa{\frac{D_s+2\bar{R}+2\lambda}{\bar{R}}}^{2b-1}}+2G^{\bar{m},s}_\phi-\frac{1}{2}G^{\bar{m},s}_t
	\Bigg\}.
\end{align}
Solving this equation is in effect the main step in producing an image, but in practice it is easier to work directly with \eqref{eq:SecondEquation}, where the constraint that $m$ and $N$ be integers can be more directly imposed.

\subsection{Image fluxes}

We now expand formula \eqref{eq:Flux} for the flux of an individual image to leading order in $\epsilon$.  Noting that
\begin{align}
	\label{eq:HatJacobians}
	\frac{\pd\lambda}{\pd\hat{\lambda}}=-\frac{1}{2M\epsilon},\qquad
	\frac{\pd q}{\pd\hat{q}}=-\frac{\sqrt{3-q^2}}{qM},\qquad
	\ab{\det
	\begin{pmatrix}
		\frac{\pd B}{\pd\hat{\lambda}} & \frac{\pd B}{\pd\hat{q}}\vspace{2pt}\\
		\frac{\pd A}{\pd\hat{\lambda}} & \frac{\pd A}{\pd\hat{q}}
	\end{pmatrix}}
	=\frac{\sqrt{3-q^2}}{2qM^2\epsilon}\ab{\det
	\begin{pmatrix}
		\frac{\pd B}{\pd\lambda} & \frac{\pd B}{\pd q}\vspace{2pt}\\
		\frac{\pd A}{\pd\lambda} & \frac{\pd A}{\pd q}
	\end{pmatrix}},
\end{align}
we have
\begin{align}
	\label{eq:FluxExpansion}
	\frac{F_o}{F_N}=\frac{\epsilon\bar{R}}{2D_s}\frac{qg^3}{\sin{\theta_o}\sqrt{1-\frac{q^2}{3}}\sqrt{\Theta_0(\theta_o)}}\ab{\det
	\begin{pmatrix}
		\frac{\pd B}{\pd\lambda} & \frac{\pd B}{\pd q}\vspace{2pt}\\
		\frac{\pd A}{\pd\lambda} & \frac{\pd A}{\pd q}
	\end{pmatrix}}^{-1},
\end{align}
where $g$ and $D_s$ are given in Eqs.~\eqref{eq:RedshiftLimit} and \eqref{eq:RadialIntegrandLimits}, respectively, and [see Eq.~\eqref{eq:AlphaBetaExpansion}]
\begin{align}
	\Theta_0(\theta_o)&=\Theta(\theta_o)|_{\lambda=0}
	=3-q^2+\cos^2{\theta_o}-4\cot^2{\theta_o}
	=\pa{\frac{\beta}{M}}^2.
\end{align}
Recall that $A$ and $B$ are defined in terms of the various integrals $I$, $J$, and $G$ by Eqs.~\eqref{eq:A} and \eqref{eq:B}.  To leading order in $\epsilon$, the determinant is
\begin{align}
	\label{eq:Determinant}
	\ab{\det
	\begin{pmatrix}
		\frac{\pd B}{\pd\lambda} & \frac{\pd B}{\pd q}\vspace{2pt}\\
		\frac{\pd A}{\pd\lambda} & \frac{\pd A}{\pd q}
	\end{pmatrix}}
	=\Bigg|&\frac{\pd}{\pd\lambda}\pa{J_r+b\tilde{J}_r}\br{\frac{\pd}{\pd q}\pa{I_r+b\tilde{I}_r}-\frac{\pd G_\theta^{m,s}}{\pd q}}\\
	&\quad-\frac{\pd}{\pd\lambda}\pa{I_r+b\tilde{I}_r}\br{\frac{\pd}{\pd q}\pa{J_r+b\tilde{J}_r}+\frac{\pd G_{t\phi}^{m,s}}{\pd q}}\Bigg|+\O{\epsilon\log{\epsilon}},\nonumber
\end{align}
where we introduced
\begin{align}
	\label{eq:RelevantG}
	G_{t\phi}^{m,s}=\frac{\hat{\lambda}G^{m,s}_\phi-\Omega_sa^2G^{m,s}_t}{M}.
\end{align}
Notice that the $\pd_\lambda G_\theta^{m,s}$ and $\pd_\lambda G_{t\phi}^{m,s}$ terms do not contribute in \eqref{eq:Determinant}; these terms are subleading on account of the factor of $\epsilon$ in the Jacobian $\pd_\lambda\hat{\lambda}$ [see Eq.~\eqref{eq:HatJacobians}].  The $\log{\epsilon}$ scaling of $m$ makes them part of the $\O{\epsilon\log{\epsilon}}$ error.

The expressions for the $G$ integrals are given at the end of App.~\ref{app:AngularIntegrals}.  The unhatted integrals scale as $\log{\epsilon}$ on account of the $\log{\epsilon}$ scaling of $m$, while the hatted integrals are $\O{\epsilon^0}$, and therefore subleading.  The derivatives of the $I$ and $J$ integrals may be computed for small $\epsilon$ by the method of matched asymptotic expansions, and the results are listed in Eqs.~\eqref{eq:RadialIntegralsVariation} of App~\ref{app:MAE}.  Using these expressions, the remaining terms in \eqref{eq:RelevantG} are given by
\begin{subequations}
\begin{align}
	\frac{\pd}{\pd\lambda}\pa{I_r+b\tilde{I}_r}&=-\frac{1}{\lambda}\br{\pa{2b-1}\frac{\bar{R}}{D_s}+\frac{1}{q}},\\
	\frac{\pd}{\pd q}\pa{I_r+b\tilde{I}_r}&=-\frac{I_r+b\tilde{I}_r}{q}+\frac{1}{q\pa{4-q^2}}\br{\pa{8-q^2}\pa{\pa{2b-1}\frac{\bar{R}}{D_s}-\frac{1}{D_o}+\frac{2}{q}}+\pa{2b-1}\frac{4\lambda}{D_s}-\frac{2R_o}{D_o}},\\
	\frac{\pd}{\pd\lambda}\pa{J_r+b\tilde{J}_r}&=\frac{1}{2\lambda}\br{\pa{2b-1}\frac{4\bar{R}+\lambda}{D_s}+\frac{7}{q}+\frac{3}{4}\frac{\pa{2b-1}D_s+q\bar{R}}{\lambda}},\\
	\frac{\pd}{\pd q}\pa{J_r+b\tilde{J}_r}&=\frac{7}{2q}\pa{I_r+b\tilde{I}_r}+\frac{1}{2}-\frac{3}{8q}\frac{\pa{2b-1}D_s+q\bar{R}}{\lambda}-\frac{11}{q^2}\\
	&\quad-\frac{1}{2q\pa{4-q^2}}\br{\pa{2b-1}\frac{\pa{32-5q^2}\bar{R}+\pa{16-q^2}\lambda}{D_s}-\frac{\pa{7-q^2}\pa{8-q^2+2R_o}}{D_o}+\frac{24}{q}}.\nonumber
\end{align}
\end{subequations}
In these expressions, we kept both the leading $\O{\log{\epsilon}}$ and subleading $\O{\epsilon}$ contributions.  It follows that the determinant \eqref{eq:Determinant} scales as $\log{\epsilon}$ with subleading $\O{\epsilon^0}$ corrections included.  Putting everything together, the flux \eqref{eq:FluxExpansion} scales as $\epsilon/\log{\epsilon}$ with subleading $\O{\epsilon}$ contributions included.  These subleading corrections are numerically important at reasonable values of $\epsilon$.

\section{Observational appearance}
\label{sec:ObservationalAppearance}

We now describe our practical approach to implementing the method described above and discuss details of the results.  We implemented this procedure in a \textsc{Mathematica} notebook, which is included with the arXiv version of this submission.

The image depends on four parameters: $\epsilon$, $\bar{R}$, $R_o$, and $\theta_o$.  One must choose $\epsilon\ll1$ and $R_o\gg1$ for our approximations to be accurate.  The radius $\bar{R}$ must satisfy $\bar{R}\geq2^{1/3}$ for the emitter to be on a stable orbit of a near-extremal black hole.  The observer inclination $\theta_o$ must satisfy $\arctan{(4/3)^{1/4}}<\theta_o<\pi/2$ for there to be any flux at all (under our approximations).  Here we will focus on the following example:
\begin{align}
	\label{eq:ExampleParameters}
	R_o=100,\qquad
	\theta_o=\frac{\pi}{2}-\frac{1}{10}=84.27^\circ,\qquad
	\epsilon=10^{-2},\qquad
	\bar{R}=\bar{R}_\mathrm{ISCO}=2^{1/3}.
\end{align}  
This describes an emitter (or hotspot) on the ISCO of a near-extremal black hole with spin $a/M=99.99995\%$, viewed from a nearly edge-on inclination.

\begin{figure}[ht!]
	\centering
	\includegraphics[width=\textwidth]{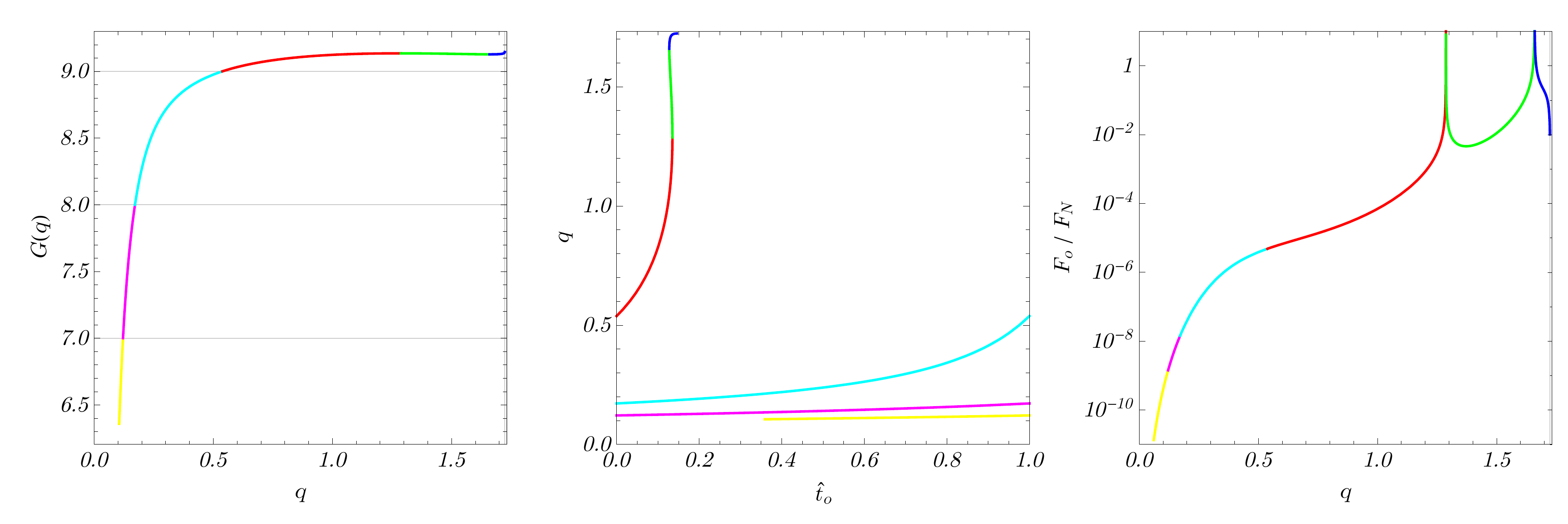}
	\caption{Determining track segments for $m=2$, $b=0$, $s=+1$ with the parameter choices of \eqref{eq:ExampleParameters}.  The condition \eqref{eq:LambdaCondition} allows the whole range of $q$.  On the left, we plot $\mathcal{G}(q)$ with color-coding explained below.  In the middle, we plot the tracks $q(t)$ of the images over the single period $0<\hat{t}_o<1$.  The yellow, magenta, and cyan curves have $N=-6,-7,-8$, respectively, and no extra label $i$, while the red, green, and blue curves have $N=-9$ and $i=1,2,3$, respectively.  The function $\mathcal{G}(q)$ actually extends to negatively infinite values near $q=0$, corresponding formally to infinitely many images (values of $N$) near the ends of the NHEKline.  However, these images are negligibly faint (rightmost plot). The complete image is formed by stitching together all track segments (labeled by $m,b,s,N,i$).}
	\label{fig:Segments}
\end{figure}

As described below Eq.~\eqref{eq:ReflectedJr}, each image is labeled by discrete parameters $m,s,b,N$ as well as an additional label if the function $\mathcal{G}(q)$ has maxima or minima.  In practice, it is easiest to choose $m,s,b$ first and then determine what range of $N$ is allowed and whether there are any extra images due to maxima or minima.  For each choice of integers $m,b,s$, the condition \eqref{eq:LambdaCondition} determines the allowed values of $q$ by the condition \eqref{eq:LambdaCondition}.  We can then determine $\lambda(q)$ [Eq.~\ref{eq:Lambda}] and hence $\mathcal{G}(q)$ [Eq.~\eqref{eq:SecondEquation}] over the allowed domain.  The range of $\mathcal{G}(q)$ over this domain determines the allowed values of $N$ by the requirement that $\hat{t}_o=\mathcal{G}+N$ lies between $0$ and $1$.  (In practice we impose a small-$q$ cutoff, as explained in Fig.~\ref{fig:Segments}.)  For each allowed value of $N$ we invert $\hat{t}_o(q)$, labeling the inverses $q_i\pa{\hat{t}_o}$ by a discrete integer $i$.  These functions $q\pa{\hat{t}_o}$ correspond to segments of an image track; each such track segment is uniquely labeled by $(m,b,s,N,i)$.  An example is shown in Fig.~\ref{fig:Segments}.

For each track segment $q\pa{\hat{t}_o}$, we may determine $\lambda\pa{\hat{t}_o}$ by \eqref{eq:Lambda}.  From these two conserved quantities, we may then compute the main observables for the segment: image position $\pa{\alpha,\beta}$ [Eq.~\eqref{eq:AlphaBetaExpansion}], image redshift $g$ [Eq.~\ref{eq:RedshiftLimit}], and image flux $F_o$ [Eq.~\eqref{eq:FluxExpansion}].  The complete observable information is built up by including all such track segments (all choices of $m,b,s,N,i$).  Formally, there are infinitely many segments since $m$ and $-N$ can become arbitrarily large, but in practice the flux is vanishingly small for all but a few values of $m$ and $N$ (see Sec.~\ref{sec:m} below and Fig.~\ref{fig:Segments} for details).  We find that the track segments line up into continuous tracks that begin and end either at the endpoints of the NHEKline with vanishing flux or as part of a pair creation/annihilation event with infinite flux (a geometrical caustic).  More generally, caustics appear when different tracks intersect.  The infinite flux can be traced to the vanishing of the derivatives $\pd\theta_s/\pd\hat{q}$ and $\pd\phi_s^\star/\pd\hat{q}$, indicating that the image extends in the vertical direction.  That is, the whole NHEKline flashes at caustics.

\begin{figure}[ht!]
	\centering
	\includegraphics[width=\textwidth]{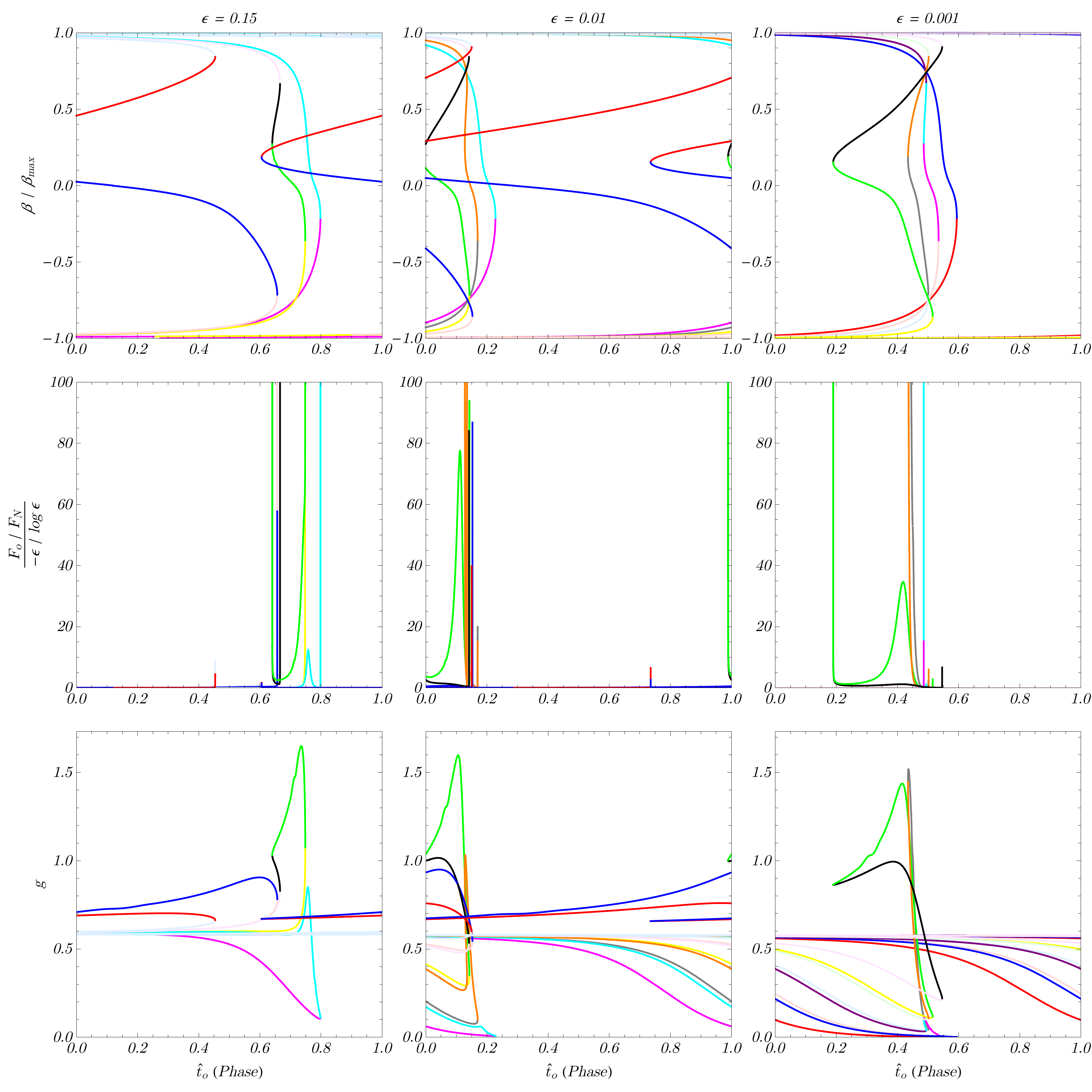}
	\caption{Positions, fluxes, and redshift factors of the brightest few images for three different values of near-extremal spin.  Left to right, we have $\epsilon=.15$ (Thorne limit), $\epsilon=.01$, and $\epsilon=.001$.  We have color-coded by continuous image tracks, each of which may be composed of multiple track segments in our accounting (different values of $m,b,N,s,i$).  For example, the primary image (green) is composed of $3$, $4$, and $5$ segments in the $\epsilon=.15$, $.01$, $.001$ cases respectively.}
	\label{fig:3x3}
\end{figure}

Figure~\ref{fig:3x3} shows the main observables for three different values of spin.  In each case, there is a bright primary image (green) together with secondary images that are important only near caustics.  The primary image is a combination of direct ($b=0$) and reflected ($b=1$) light, with the transition occurring near peak flux.  These photons are emitted near the forward direction (equivalently $g$ near $\sqrt{3}$) and orbit the black hole $\O{\epsilon^{-1}}$ times, while crossing the equatorial plane $\O{\log{\epsilon}}$ times, before emerging from the throat.  For example, at $\epsilon=.01$, the primary image is composed of segments with two and three equatorial crossings ($i.e.$, $m=2$ and $3$), and with winding number around the axis of symmetry [Eq.~\eqref{eq:DeltaPhiLimit}] ranging between $17$ and $23$.  The peak redshift factor of $g\approx1.6$ corresponds to light emitted in a cone of $27^\circ$ around the forward direction.  For the secondary images, we have included a representative selection to illustrate the structure.  The typical redshift in this case is $g=1/\sqrt{3}$ (corresponding to $\lambda\sim0$).  As the spin is increased, the typical position and redshift of the images do not change, while the typical flux scales as $\epsilon/\log{\epsilon}$.

We see that the light emerges with typical redshift factor of either $g=\sqrt{3}$ (blueshifted primary image) or $g=1/\sqrt{3}$ (redshifted secondary images).  These factors will shift the iron line at $E_{\mathrm{FeK}\alpha}=6.4$~keV to $11.1$~keV and $3.7$~keV, respectively.  Astronomical observation of spectral lines at such frequencies could be an indication of high-spin black holes.  This is tantalizingly close to the observed peak at 3.5 keV \cite{Carlson2015}.

\subsection{Peak flux}
\label{sec:m}

\begin{figure}[ht!]
	\centering
	\includegraphics[width=\textwidth]{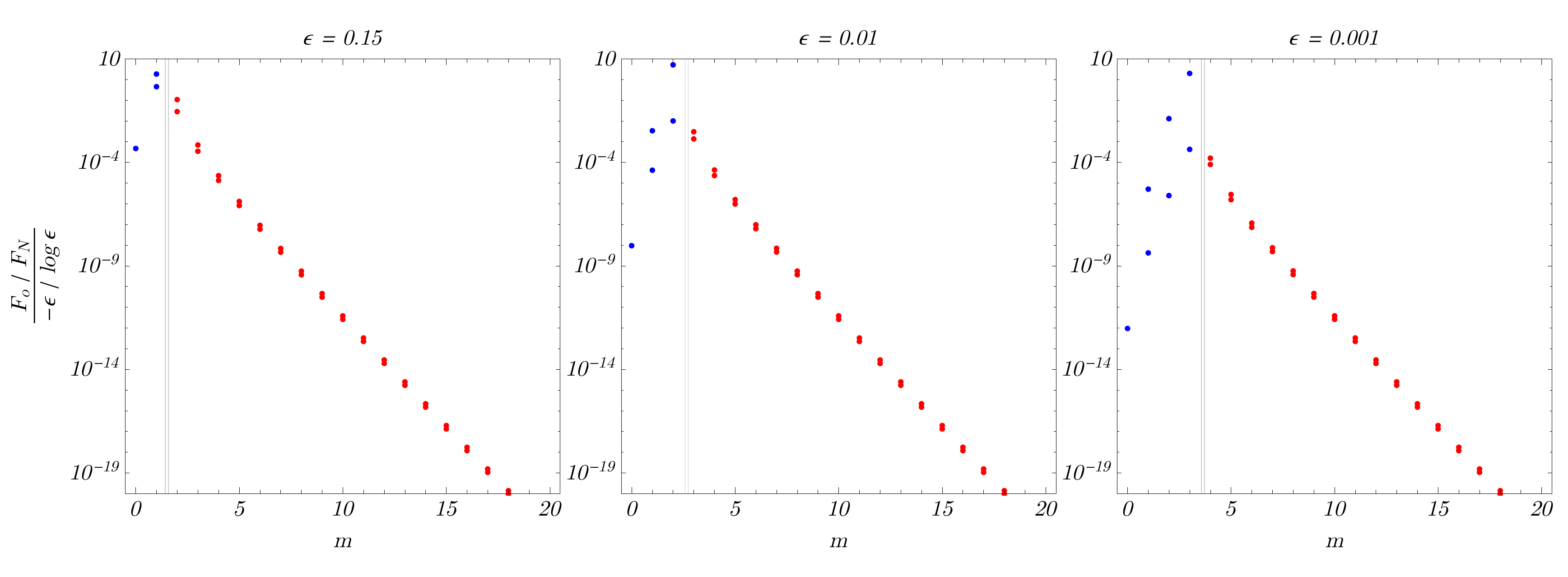}
	\caption{Left to right: plots of $F_o/F_N$ for $\epsilon=.15$ (Thorne limit), $\epsilon=.01$, and $\epsilon=.001$, with parameters otherwise as in Eq.~\eqref{eq:ExampleParameters}.  We set $q=3/2$ and let $m$ vary.  Blue and red correspond to direct ($b=0$) and reflected ($b=1$), respectively.  At each value of $m$ and $b$, there are two images, corresponding to $s=\pm1$, except for the special case $m=0$, which has only one sign of $s$.  The vertical lines are at the predicted peak value of flux, $m=m_0$ for each value of $s\in\cu{1,-1}$.}
	\label{fig:mFlux}
\end{figure}

\begin{figure}[ht!]
	\centering
	\includegraphics[width=.4\textwidth]{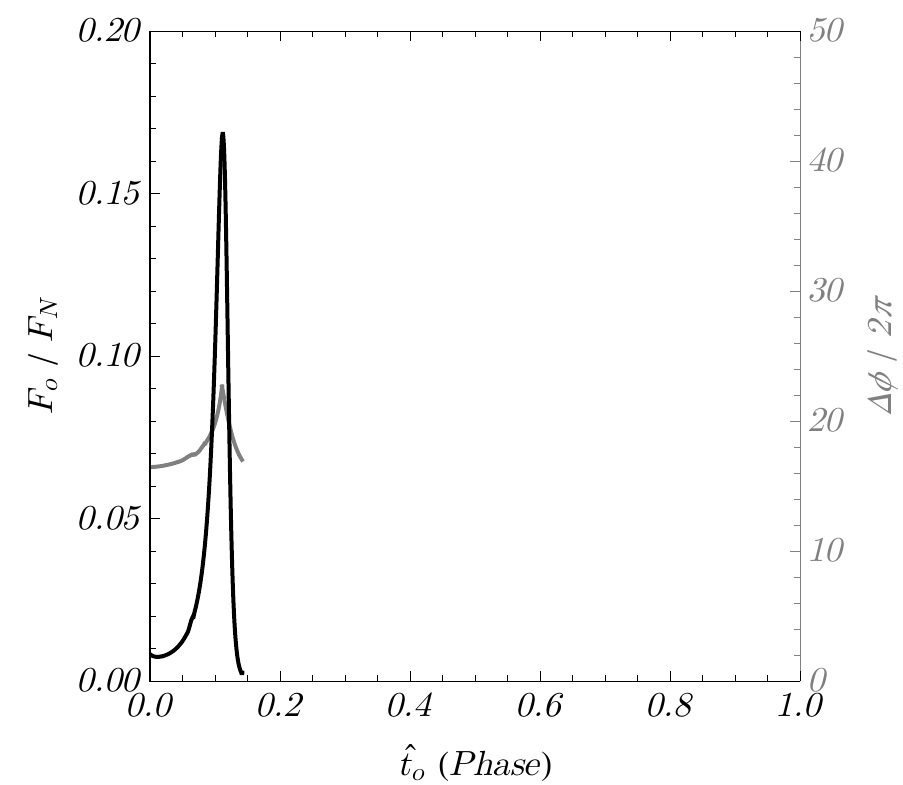}
	\caption{The flux (black) and winding number (gray) of the primary image for the parameters \eqref{eq:ExampleParameters}.  The position of the source at the time of emission is given by $\mod_{2\pi}\Delta\phi$.  Notice that the single peak in observed flux corresponds to several orbits of the source.}
	\label{fig:Winding}
\end{figure}

The complete image is assembled from all the track segments $(m,b,s,N,i)$.  The parameters $b,s,N,i$ have finite ranges, while in principle $m$ can take any value $m\geq0$.  However, according to the discussion surrounding Eq.~\eqref{eq:m}, we expect that only values of $m\sim-1/(qG_\theta)\log{\epsilon}$ should matter.  Our numerical analysis confirms this suspicion and reveals that the precise value of the maximum flux is the special value $m_0$ such that the radius of emission $\bar{R}$ precisely coincides with the radial turning point $x_\mathrm{min}$ [Eq.~\eqref{eq:TurningRadius}] of the photon trajectory.  This is determined by plugging $\bar{R}=x_\mathrm{min}$ (equivalently $D_s=0$) into the geodesic equation \eqref{eq:FirstEquation},
\begin{align}
    I_r|_{\bar{R}=x_\mathrm{min}}=m_0G_\theta-s\hat{G}_\theta.
\end{align}
Note that the dependence on $b$ has dropped out because $\tilde{I}_r$ vanishes when $D_s=0$.  Explicitly, we have
\begin{align}
    \label{eq:mPeak}
	m_0=s\frac{\hat{G}_\theta}{G_\theta}+\frac{1}{qG_\theta}\log\br{\frac{4q^2}{q^2+qD_o+2R_o}\pa{1+\frac{2}{\sqrt{4-q^2}}}\frac{R_o}{\epsilon\bar{R}}}.
\end{align}
This special value of $m$ also corresponds to the boundary between direct ($b=0$) and reflected ($b=1$) light.  Indeed, using $e^{qG_\theta^{\bar{m},s}}=\epsilon e^{qG_\theta^{m,s}}$, the conditions \eqref{eq:LambdaCondition} can be expressed as
\begin{subequations}
\begin{align}
	m&<m_0\qquad\text{if $b=0$ (direct)},\\
	m&>m_0\qquad\text{if $b=1$ (reflected)}.
\end{align}
\end{subequations}
Figures~\ref{fig:mFlux} and \ref{fig:Winding} demonstrate the peak flux and its properties.  In Fig.~\ref{fig:mFlux}, we show the dependence of flux on $m$ at fixed $q$, showing the peak at $m=m_0$ and exponential falloff on either side.  In Fig.~\ref{fig:Winding}, we show the peak flux in the time domain along with the winding number $\Delta\phi/2\pi$.  The cusp in the winding number is associated with $D_s=0$ [see Eq.~\eqref{eq:DeltaPhiLimit}], showing that peak flux is indeed $D_s=0$ ($i.e.$, $m=m_0$).

Finally, we derive simple formulae for the flux, redshift factor, and winding number at peak flux.  Since $m$ must be an integer, $m=m_0$ forces $q$ to take specific values found by solving Eq.~\eqref{eq:mPeak}.  Of the several such solutions for $q$, we find numerically that the peak flux corresponds to the largest value of $q$ in the allowed range \eqref{eq:NHEKline}.  (We also find that as $\epsilon\to0$, the other solutions for $q$ correspond to the fluxes of secondary images at the same moment.)  When $m=m_0$ and $\bar{R}=x_\mathrm{min}$, we have
\begin{align}
    \label{eq:BrightestImage}
    \lambda=-P\bar{R},\qquad
    \Upsilon=(1-P)P\bar{R},\qquad
    P\equiv1-\sqrt{1-\frac{q^2}{4}}\ge0,
\end{align}
We may now plug $m=m_0$ ($i.e.$, Eq.~\eqref{eq:BrightestImage} and therefore $D_s=0$) into Eqs.~\eqref{eq:RedshiftLimit}, \eqref{eq:DeltaPhiLimit}, and \eqref{eq:FluxExpansion} to determine the flux, redshift, and winding number at the moment of peak flux. We find
\begin{subequations}
\begin{align}
    \label{eq:Peak}
    &\frac{F_o}{F_N}=\frac{\epsilon\bar{R}}{X}\frac{qg^3}{\sin{\theta_o}\sqrt{1-\frac{q^2}{3}}\sqrt{\Theta_0(\theta_o)}},\qquad
    g=\frac{1}{\sqrt{3}-\frac{4}{\sqrt{3}}P},\qquad
	\Delta\phi=\frac{q}{2P}\frac{1}{\epsilon\bar{R}},\\
	X&\equiv2D_s\ab{\det
	\begin{pmatrix}
		\frac{\pd B}{\pd\lambda} & \frac{\pd B}{\pd q}\vspace{2pt}\\
		\frac{\pd A}{\pd\lambda} & \frac{\pd A}{\pd q}
	\end{pmatrix}}\\
	&=\ab{\pa{1-\frac{4}{P}}\frac{\pd G_\theta^{m,s}}{\pd q}+\frac{2}{P}\frac{\pd G_{t\phi}^{m,s}}{\pd q}+\frac{1+\frac{3-q^2}{P}}{4-q^2}\pa{1-\frac{q^2-2R_o-8}{qD_o}-\frac{16}{q^2}}+\pa{1+\frac{3}{P}}\frac{I_r|_{\bar{R}=x_\mathrm{min}}}{q}}.
\end{align}
\end{subequations}
Note that the dependence on the parameter $b\in\cu{0,1}$, which is undefined at the transition between direct and reflected light, has again dropped out, this time because it only entered through an overall factor of $\ab{2b-1}=1$ multiplying $X$.  We see from Eq.~\eqref{eq:Peak} that the redshift factor at peak flux is bounded below by $1/\sqrt{3}$, so a redshift factor of at least $1/\sqrt{3}$ will always be reached by at least one image over each period.  Numerically, we also find that the redshift factor is always maximized at peak flux, so Eq.~\eqref{eq:Peak} gives an \textit{upper} bound on the redshift factor over the image period.

To summarize: The peak flux, as well as the associated redshift factor and winding number, are given by Eq.~\eqref{eq:Peak}, using the largest value of $q$ [in the allowed range \eqref{eq:NHEKline}] such that $m_0$ [Eq.~\eqref{eq:mPeak}] is an integer.

\subsection*{Acknowledgements}

This work was supported in part by NSF grants 1205550 to Harvard University and 1506027 to the University of Arizona.  S.G. thanks Dimitrios Psaltis and Feryal \"Ozel for helpful conversations.  A.L. is grateful to Sheperd Doeleman, Michael D. Johnson, Achilleas Porfyriadis, and Yichen Shi for fruitful discussions.  Many of these took place at the Black Hole Initiative at Harvard University, which is supported by a grant from the John Templeton Foundation.
\hfill\includegraphics[scale=.02]{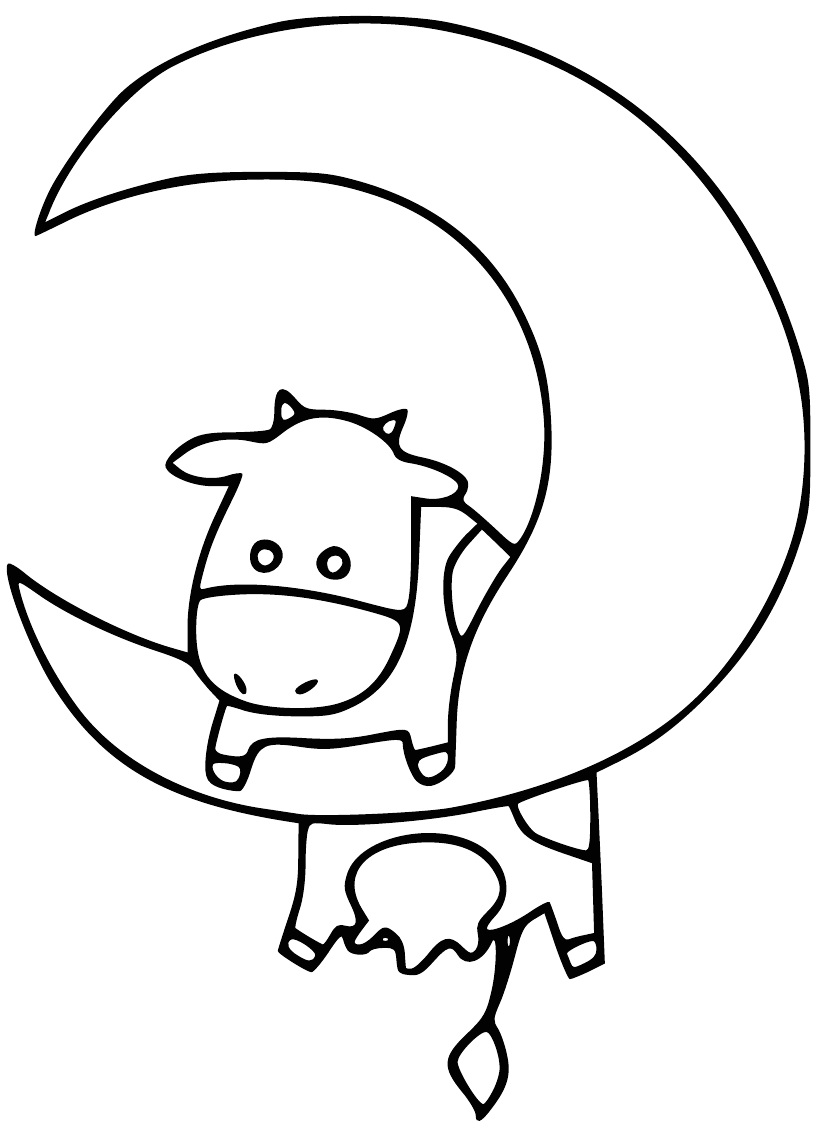}

\appendix

\section{Shadow and NHEKline}
\label{app:Shadow}

A useful reference for understanding images from light near a black hole is the black hole ``shadow,'' defined as the portion of the image that would be dark if the black hole were uniformly backlit \cite{Bardeen1973}.  We now briefly review the basics and present the extremal limit.  The edge of the shadow is set by the threshold between captured and escaping photons, which corresponds to unstable massless geodesics with fixed $r=\tilde{r}$.  These ``spherical photon orbits'' satisfy
\begin{align}
	\mathcal{R}\pa{\tilde{r}}=\mathcal{R}'\pa{\tilde{r}}=0,
\end{align}
where prime denotes derivative.  Provided $0<a<M$, these equations imply
\begin{align}
	\label{eq:Shadow}
	\hat{\lambda}=-\frac{\tilde{r}^2\pa{\tilde{r}-3M}+a^2\pa{\tilde{r}+M}}{a\pa{\tilde{r}-M}},\qquad
	\hat{q}=\frac{\tilde{r}^{3/2}}{a\pa{\tilde{r}-M}}\sqrt{4a^2M-\tilde{r}\pa{\tilde{r}-3M}^2}.
\end{align}
As noted following \eqref{eq:ConservedQuantities}, orbits crossing the equatorial plane have real $\hat{q}$.  Thus there is a spherical photon orbit only when the quantity under the square root \eqref{eq:Shadow} is positive.  This entails a condition on the radius $\tilde{r}$:
\begin{align}
	\label{eq:ShadowParameterization}
	\tilde{r}\in\br{\tilde{r}_-,\tilde{r}_+},\qquad
	\tilde{r}_\pm\equiv2M\br{1+\cos\pa{\frac{2}{3}\arccos{\pm\frac{a}{M}}}}.
\end{align}
The shadow edge is the curve $\pa{\alpha\pa{\tilde{r}},\beta\pa{\tilde{r}}}$ obtained by plugging Eqs.~\eqref{eq:Shadow} for $\hat{q}$ and $\hat{\lambda}$ into Eq.~\eqref{eq:ScreenCoordinates}.  The parameter $\tilde{r}$ ranges over the region where \eqref{eq:ShadowParameterization} is satisfied and $\beta$ is real.  (The latter condition is equivalent to the requirement that photons near to the unstable orbit can reach asymptotic infinity at the desired inclination.)  It is possible to express the precise range of $\tilde{r}$ in closed form as the solution to a quartic, but the expression is not particularly helpful.

\subsection{Extremal limit}

The extremal limit of the shadow is slightly subtle.  Letting $a\to M$ in Eqs.~\eqref{eq:Shadow} and \eqref{eq:ShadowParameterization} gives
\begin{subequations}
\begin{align}
	M\hat{\lambda}&=-\pa{\tilde{r}^2-M^2-2M\tilde{r}},\\
	M\hat{q}&=\tilde{r}^{3/2}\sqrt{4M-\tilde{r}},\\
	\pa{\tilde{r}_-,\tilde{r}_+}&=(M,4M).
\end{align}
\end{subequations}
The shadow edge is then the curve traced by
\begin{subequations}
\label{eq:FarShadow}
\begin{align}
	M\alpha\pa{\tilde{r}}&=\pa{\tilde{r}^2-M^2-2M\tilde{r}}\csc{\theta_o},\\
	M\beta\pa{\tilde{r}}&=\pm\sqrt{\tilde{r}^3\pa{4M-\tilde{r}}+M^4\cos^2{\theta_o}-\pa{\tilde{r}^2-M^2-2M\tilde{r}}^2\cot^2{\theta_o}}.
\end{align}
\end{subequations}
However, this curve is \textit{not} in general closed: it takes the shape shown in the dotted line in Fig.~\ref{fig:OpticalAppearance}, with the two endpoints at
\begin{align}
	\pa{\alpha_\mathrm{end},\beta_\mathrm{end}}=\pa{\alpha,\beta}|_{\tilde{r}=M}
	=\pa{-2M\csc\theta_o,\pm M\sqrt{3+\cos^2{\theta_o}-4\cot^2{\theta_o}}}.
\end{align}
When the quantity under the square root is negative, there are no endpoints and the curve is closed.  The condition for an open curve is thus $\theta_\mathrm{crit}<\theta_o<\pi-\theta_\mathrm{crit}$, where
\begin{align}
	\theta_\mathrm{crit}=\arctan\br{(4/3)^{1/4}}
	\approx47^\circ.
\end{align}

Since the curve does close for all $a<M$, our $a\to M$ limit has missed an important piece.  This piece comes from parameter values $\tilde{r}$ arbitrarily near $M$.  To recover it, we use an alternative parameter $\tilde{R}$ defined by
\begin{align}
	a=M\sqrt{1-\delta^2},\qquad
	\tilde{r}=M\pa{1+\delta\tilde{R}}.
\end{align}
Now Eqs.~\eqref{eq:Shadow} and \eqref{eq:ShadowParameterization} give
\begin{subequations}
\begin{align}
	\label{eq:LambdaShadow}
	\hat{\lambda}&=2M+\O{\delta},\\
	\label{eq:qShadow}
	\hat{q}&=M\sqrt{3-\frac{4}{\tilde{R}^2}}+\O{\delta},\\
	\tilde{r}_-&=M\pa{1+\frac{2}{\sqrt{3}}\delta}+\O{\delta^2},\\
	\tilde{r}_+&=4M+\O{\delta^2},
\end{align}
\end{subequations}
and now the shadow edge is the curve traced by
\begin{subequations}
\begin{align}
	M\alpha\pa{\tilde{R}}&=-2M^2\csc{\theta_o}+\O{\delta},\\
	M\beta\pa{\tilde{R}}&=\pm M^2\sqrt{3+\cos^2{\theta_o}-4\cot^2{\theta_o}-\frac{4}{\tilde{R}^2}}+\O{\delta}.
\end{align}
\end{subequations}
The allowed range of $\tilde{R}$ can now be determined from the twin requirements that $\tilde{r}\in\br{\tilde{r}_-,\tilde{r}_+}$ and $\beta\in\mathbb{R}$.  The result is
\begin{align}
	\label{eq:NearShadowParameter}
	\tilde{R}\in\left[\frac{2}{\sqrt{3+\cos^2{\theta_o}-4\cot^2{\theta_o}}},\infty\right)+\O{\delta}.
\end{align}
As $\delta\to0$, this precisely covers the missing line in Fig.~\ref{fig:OpticalAppearance}.  We name this segment the NHEKline in light of the fact that emission near the line originates from the NHEK region of the spacetime (App.~\ref{app:NHEK} below).

We can determine the ranges of the conserved quantities $\hat{\lambda}$ and $\hat{q}$ by plugging Eq.~\eqref{eq:NearShadowParameter} into Eqs.~\eqref{eq:LambdaShadow}--\eqref{eq:qShadow}, and from these we can obtain the ranges of the alternative $\lambda$ and $q$.  Putting everything together, the NHEKline can be characterized in three equivalent ways:
\begin{subequations}
\label{eq:NHEKline}
\begin{align}
	\text{Screen coordinates}:&\qquad
	\alpha=-2M\csc{\theta_o},\qquad
	\ab{\beta}<M\sqrt{3+\cos^2{\theta_o}-4\cot^2{\theta_o}},\\
	\text{Conserved quantities }\pa{\hat{\lambda},\hat{q}}:&\qquad
	\hat{\lambda}=2M,\qquad
	M\cos{\theta_o}\sqrt{3+4\cot^2{\theta_o}}
	=\hat{q}_\mathrm{min}<\hat{q}<\sqrt{3}M,\\
	\text{Conserved quantities }(\lambda,q):&\qquad
	\lambda=0,\qquad
	0<q<q_\mathrm{max}=\sqrt{3+\cos^2{\theta_o}-4\cot^2{\theta_o}}.
\end{align}
\end{subequations}
This line exists provided $\theta_\mathrm{crit}<\theta_o<\pi-\theta_\mathrm{crit}$.

To summarize: the extremal limit of the Kerr shadow is given by the union of the curve \eqref{eq:FarShadow} and the NHEKline \eqref{eq:NHEKline}.  This reflects two separate extremal limits of the Kerr spacetime.

\section{NHEK sources}
\label{app:NHEK}

In the presentation given in the main body, the distinctive feature of the calculation---all light appearing on the NHEKline---simply emerged from the detailed calculations.  We now explain the spacetime-geometrical origin of this feature and thereby establish that it persists for near-horizon sources more generally.  The key observation is that source and observer are adapted to two distinct extremal limits that have a singular relationship, forcing generic light from the source to appear in a special location on the observer screen.  

The variety of extremal limits and their physical consequences are discussed in Refs.~\cite{Gralla2015,Porfyriadis2014b,Gralla2016a,Gralla2016c}.  We can discuss the different limits in terms of barred coordinates defined by
\begin{align}
	\tilde{T}=\frac{\kappa^pt}{2M},\qquad
	\tilde{R}=\frac{R}{\kappa^p},\qquad
	\tilde{\Theta}=\theta,\qquad
	\tilde{\Phi}=\phi-\frac{t}{2M},\qquad
	a=M\sqrt{1-\kappa^2}.
\end{align}
Letting $\kappa\to0$ at fixed barred coordinates, different choices of $p$ give rise to different limits.  The choice $p=0$ is the usual limit to the extremal Kerr exterior, while any choice $0<p\leq 1$ gives a patch of the NHEK metric.  When $0<p<1$ it is the Poincar\'e patch, while the special case $p=1$ gives a smaller, ``near-NHEK'' patch.  For $p>1$ the limit is singular.

When a tensor field has a finite, non-zero extremal limit with $p>0$, we say that it has a good near-horizon limit.  We will show that any photon whose four-momentum $p^\mu$ has a good near-horizon limit arrives on the NHEKline.  First, note that
\begin{align}
	\label{eq:KerrIsometries}
	\pd_\phi=\pd_{\tilde{\Phi}},\qquad
	\pd_t+\frac{1}{2M}\pd_\phi=\frac{\kappa^p}{2M}\pd_{\tilde{T}}.
\end{align}
Now fix a value $p>0$.  A particle whose four-momentum $p^\mu$ has a good near-horizon limit ($i.e.$, finite extremal limit with $p>0$) will have a finite ``energy'' $-p_\mu\pd_{\tilde{T}}^\mu$.  However, given $p>0$, Eq.~\eqref{eq:KerrIsometries} shows that the usual energy $E=-p_\mu\pd_t^\mu$ and angular momentum $L=p_\mu\pd_\phi^\mu$ will satisfy $\hat{\lambda}=2M$ as $\kappa\to0$.  (Recall the definition $\hat{\lambda}=L/E$.)  This reproduces Eq.~\eqref{eq:LambdaExpansion} as $\epsilon\to0$, and the same arguments then given in the text prove that the photon ends up on the NHEKline.  Similar arguments were given by Bardeen \cite{Bardeen1973} in the special case of an extremal black hole and an equatorial observer and in Ref.~\cite{Porfyriadis2016} more generally.

Our observer sits at infinity and hence is adapated to the usual extremal limit ($p=0$), which preserves the asymptotically flat region.  Mathematically, the observer four-velocity $u_{o}^\mu$ has a finite, non-zero extremal limit when $p=0$.  On the other hand, our source is on/near the ISCO and hence is adapted to $p=2/3$ [see Eqs.~\eqref{eq:aExpansion} and \eqref{eq:rExpansion}].  Mathematically, the source four-velocity $u_{s}^\mu$ has a finite, non-zero extremal limit when $p=2/3$.  Any photons emitted by the source will similarly have four-momenta $p^\mu$ with finite, non-zero extremal limits when $p=2/3$, and hence arrive on the NHEKline.  Thus the appearance of the light on the NHEKline can be attributed to the fact that we consider a near-horizon source, the $p=2/3$ ISCO.

The ISCO is not the only physically interesting near-horizon orbit.  For example, the innermost bound circular orbit (IBCO) has a finite limit with $p=1$ (see, $e.g.$, Ref.~\cite{Gralla2016c}), and there are also out-of-equatorial orbits that are similarly adapted to near-horizon limits.  Most models of accretion disks terminate with orbits lying somewhere between the ISCO and IBCO \cite{Abramowicz2013}.  The plunging portion of the accretion flow is slightly perturbed from the inner edge of the disk and hence will also be near-horizon adapted.  (In fact, it should be possible to calculate the flux from a plunging emitter by mapping our circular-orbit calculation using the techniques of Ref.~\cite{Porfyriadis2014b}.)  In summary, a wide variety of physical processes will generate $p>0$ photons, which therefore end up on the NHEKline.  We also expect that the general features of the light curves and redshifts seen in our calculation will persist more generally, as these are determined essentially by the near-horizon geometry shared by more general sources.

\section{Radial integrals}
\label{app:MAE}

We now describe the computation of the various radial integrals in the limit $\epsilon\to0$.  This generalizes the extremal case computations in Ref.~\cite{Porfyriadis2016}.  We discuss two representative examples in detail and then quote results for the remainder.

\subsection{Matched asymptotic expansion example: \texorpdfstring{$I_r$}{direct Ir}}

We are interested in computing the integral
\begin{align}
	\label{eq:IrComputation}
	I_r=M\int_{r_s}^{r_o}\frac{\ed r}{\sqrt{\mathcal{R}(r)}},\qquad
	\mathcal{R}(r)&=\pa{r^2+a^2-a\hat{\lambda}}^2-\pa{r^2-2Mr+a^2}\br{\hat{q}^2+\pa{a-\hat{\lambda}}^2}.
\end{align}
This is an elliptic integral at any value of spin $a$, but the closed-form expression is intractable because it requires the roots of the quartic polynomial $\mathcal{R}(r)$.  However, the integral may be approximated using matched asymptotic expansions under the regime of relevance to the paper.  This regime is $\epsilon\to0$ with [copying Eqs.~\eqref{eq:aExpansion}, \eqref{eq:rExpansion}, \eqref{eq:LambdaExpansion}, \eqref{eq:qExpansion}]
\begin{align}
	\label{eq:NearExtremalExpansion}
	a=M\sqrt{1-\epsilon^3},\qquad
	r_s=M\pa{1+\epsilon\bar{R}},\qquad
	\hat{\lambda}=2M\pa{1-\epsilon\lambda},\qquad
	\hat{q}=M\sqrt{3-q^2}.
\end{align}
We may equivalently use Eq.~\eqref{eq:ExtremalExpansion}.

We will use the radial coordinate $R$ given in Eq.~\eqref{eq:R}.  We introduce constants $0<p<1$ and $C>0$ and split the integral into two pieces,
\begin{align}
	I_r=M^2\int_{\epsilon\bar{R}}^{\epsilon^pC}\frac{\ed R}{\sqrt{\mathcal{R}}}+M^2\int_{\epsilon^pC}^{R_o}\frac{\ed R}{\sqrt{\mathcal{R}}}.
\end{align}
There is no approximation at this stage.  However, the scaling of $\epsilon^p$ (with $0<p<1$)  introduces a separation of scales $\epsilon\ll\epsilon^{p}\ll1$ as $\epsilon\to0$.  Given the limits of integration, we may approximate the first integral using $R\sim\epsilon$ and the second integral using $R\sim1$.  The constants $p$ and $C$ will of course cancel out of the final answer.

To compute the first integral, we make the change of variables $x=R/\epsilon$ and expand in $\epsilon$ at fixed $x$.  The answer is:
\begin{align}
	\label{eq:NearIntegrand}
	M^2\int_{\epsilon\bar{R}}^{\epsilon^pC}\frac{\ed R}{\sqrt{\mathcal{R}}}&= M^2\int_{\bar{R}}^{\epsilon^{p-1}C}\pa{\frac{1}{\sqrt{q^2x^2+4\lambda\pa{2x+\lambda}}}+\O{\epsilon}}\ed x\\
	\label{eq:NearIntegral}
	&=\frac{1}{q}\br{\log\pa{\frac{2q^2}{q^2\bar{R}+qD_s+4\lambda}}-\frac{1}{2}\log{\epsilon}+\log{C}}+\O{\epsilon^p},
\end{align}
where $D_s$ was defined in \eqref{eq:RadialIntegrandLimits}.  We determined the $\O{\epsilon^p}$ error scaling in \eqref{eq:NearIntegral} by keeping subleading terms in \eqref{eq:NearIntegrand}.  To compute the second integral we expand in $\epsilon$ at fixed $R$. The answer is:
\begin{align}
	\label{eq:FarIntegrand}
	M^2\int_{\epsilon^p C}^{R_o}\frac{\ed R}{\sqrt{\mathcal{R}}}
	&=M^2\int_{\epsilon^p C}^{R_o}\pa{\frac{\ed R}{R\sqrt{q^2+4R+R^2}}+\O{\epsilon}}\\
	\label{eq:FarIntegral}
	&=\frac{1}{q}\br{\log\pa{\frac{2q^2R_o}{q^2+qD_o+2R_o}}-\frac{1}{2}\log\epsilon-\log{C}}+\O{\epsilon^p},
\end{align}
where $D_o$ was defined in \eqref{eq:RadialIntegrandLimits}. 
Again, the error scaling in \eqref{eq:FarIntegral} was determined by keeping subleading terms in \eqref{eq:FarIntegrand}.  Adding Eqs.~\eqref{eq:NearIntegral} and \eqref{eq:FarIntegral} gives the complete integral:
\begin{align}
	\label{eq:MatchingIntegrals}
	I_r=-\frac{1}{q}\log{\epsilon}+\frac{1}{q}\log\br{\frac{4q^4R_o}{\pa{q^2+qD_o+2R_o}\pa{q^2\bar{R}+qD_s+4\lambda}}}+\O{\epsilon}.
\end{align}
Once again, the $\O{\epsilon}$ scaling of the error term was determined by keeping subleading terms ($i.e.$, performing the matched asymptotic expansion to the next non-trivial order).  The coefficients of $\O{\epsilon^p}$ terms all properly cancel out, leaving corrections scaling like $\epsilon$.  We have verified \eqref{eq:MatchingIntegrals} numerically against the exact integral \eqref{eq:IrComputation}.

\subsection{Second example: \texorpdfstring{$\tilde{I}_r$}{reflected Ir}}
\label{app:ReflectedRadialIntegral}

We are also interested in computing the integral
\begin{align}
	\tilde{I}_r=2M\int_{r_\mathrm{min}}^{r_s}\frac{\ed r}{\sqrt{\mathcal{R}(r)}},\qquad
	\mathcal{R}(r)&=\pa{r^2+a^2-a\hat{\lambda}}^2-\pa{r^2-2Mr+a^2}\br{\hat{q}^2+\pa{a-\hat{\lambda}}^2}.
\end{align}
Again, the ``exact'' expression is intractable and we instead compute as $\epsilon\to0$ given \eqref{eq:NearExtremalExpansion}.  Here $r_\mathrm{min}$ is the largest root of $\mathcal{R}(r)$ such that $r_\mathrm{min}<r_s$.  The requirement $r_\mathrm{min}<r_s$ means that $r_\mathrm{min}$ must (like $r_s$) also approach the horizon as $\epsilon\to0$.  We find that there are only roots for $R\sim\epsilon$ (rather than, say $R\sim\epsilon^{3/2}$).  This means that the integral takes place entirely in the NHEK region $R\sim\epsilon$ (as opposed to having a piece in the near-NHEK region $R\sim\epsilon^{3/2}$).   Introducing $x=R/\epsilon$, we have 
\begin{align}
	\sqrt{\mathcal{R}}=\epsilon M^2\sqrt{q^2x^2+4\lambda\pa{2x+\lambda}}+\O{\epsilon^2}.
\end{align}
The larger root is
\begin{align}
	\label{eq:TurningRadius}
	x_\mathrm{min}=\frac{2}{q^2}\pa{-2\lambda+\ab{\lambda}\sqrt{4-q^2}}+\O{\epsilon}.
\end{align}  
(For $\lambda>0$, the turning point is inside the horizon, but we may still compute the integral.  The non-existence of $\tilde{J}_r$ precludes the existence of a valid geodesic trajectory.)  We may now compute the integral as
\begin{subequations}
\begin{align}
	\label{eq:ReflectedIrComputation}
	\tilde{I}_r&=2M^2\int_{x_\mathrm{min}}^{\bar{R}}\pa{\frac{\ed x}{\sqrt{q^2x^2+4\lambda\pa{2x+\lambda}}}+\O{\epsilon}}\\
	&=\frac{1}{q}\log\br{\frac{\pa{q^2\bar{R}+qD_s+4\lambda}^2}{4\pa{4-q^2}\lambda^2}}+\O{\epsilon}.
\end{align}
\end{subequations}
We have determined the $\O{\epsilon}$ scaling of the errors by keeping subleading terms in \eqref{eq:TurningRadius} and \eqref{eq:ReflectedIrComputation}.

\subsection{List of results}

We now list the results of using these methods to compute the radial integrals that appear in the Kerr lens equations~\eqref{eq:LensEquations} in the regime \eqref{eq:NearExtremalExpansion}.  We find:
\begin{subequations}
\begin{align}
	I_r&=M\int_{r_s}^{r_o}\frac{\ed r}{\sqrt{\mathcal{R}(r)}}
	=-\frac{1}{q}\log{\epsilon}+\frac{1}{q}\log\br{\frac{4q^4R_o}{\pa{q^2+qD_o+2R_o}\pa{q^2\bar{R}+qD_s+4\lambda}}}+\O{\epsilon},\\
	\tilde{I}_r&=2M\int_{r_\mathrm{min}}^{r_s}\frac{\ed r}{\sqrt{\mathcal{R}(r)}}
	=\frac{1}{q}\log\br{\frac{\pa{q^2\bar{R}+qD_s+4\lambda}^2}{4\pa{4-q^2}\lambda^2}}+\O{\epsilon},\\
	J_r&=\int_{r_s}^{r_o}\frac{\mathcal{J}_r}{\sqrt{\mathcal{R}(r)}}\ed r\qquad\qquad\qquad\qquad
	\mathcal{J}_r=\frac{a\pa{2Mr-a\hat{\lambda}}-\Omega_sr\br{r^3+a^2\pa{r+2M}-2aM\hat{\lambda}}}{\Delta}\\
	&=-\frac{7}{2}I_r+\frac{q}{2}\pa{1-\frac{3}{4}\frac{\bar{R}}{\lambda}}-\frac{1}{2}\pa{D_o-\frac{3}{4}\frac{D_s}{\lambda}}+\log\br{\frac{\pa{q+2}^2\bar{R}}{\pa{D_o+R_o+2}\pa{D_s+2\bar{R}+2\lambda}}}+\O{\epsilon},\\
	\tilde{J}_r&=2\int^{r_s}_{r_\mathrm{min}}\frac{\mathcal{J}_r}{\sqrt{\mathcal{R}(r)}}\ed r
	=-\frac{7}{2}\tilde{I}_r-\frac{3}{4}\frac{D_s}{\lambda}+\log\br{\frac{\pa{D_s+2\bar{R}+2\lambda}^2}{\pa{4-q^2}\bar{R}^2}}+\O{\epsilon},
\end{align}
\end{subequations}
where $D_s$ and $D_o$ are as defined in Eq.~\eqref{eq:RadialIntegrandLimits},
\begin{align}
	D_s=\sqrt{q^2\bar{R}^2+8\lambda\bar{R}+4\lambda^2},\qquad
	D_o=\sqrt{q^2+4R_o+R_o^2}.
\end{align}
These results appear in the text in Eqs.~\eqref{eq:Ir}, \eqref{eq:Jr} and \eqref{eq:ReflectedJr}.  Next, the variations are\footnote{In the course of our computation, we encountered errors in Ref.~\cite{Cunningham1973}.  Equations~(A13) \& (A15) are missing a factor of $2$ in the term $\sqrt{r_s^2+r_s^{3/2}}$, which should instead read $\sqrt{r_s^2+2r_s^{3/2}}$. This factor correctly appears in Eq.~(A5) but is subsequently dropped.  The last term in Eq.~(A20) displays the integrand $r^4-\hat{q}^2\pa{r-1}^2$ instead of the correct $r^4-\hat{q}^2r\pa{r-2}$.  In the (near-)extremal limit, this changes the $\epsilon$-scaling and therefore qualitatively alters the result.}
\begin{subequations}
\label{eq:RadialIntegralsVariation}
\begin{align}
	\frac{\pd I_r}{\pd\lambda}&=\frac{1}{\lambda}\pa{\frac{\bar{R}}{D_s}-\frac{1}{q}},\\
	\frac{\pd\tilde{I}_r}{\pd\lambda}&=-\frac{2}{\lambda}\frac{\bar{R}}{D_s},\\
	\frac{\pd I_r}{\pd q}&=-\frac{I_r}{q}-\frac{1}{q\pa{4-q^2}}\br{\pa{8-q^2}\pa{\frac{\bar{R}}{D_s}+\frac{1}{D_o}-\frac{2}{q}}+\frac{4\lambda}{D_s}+\frac{2R_o}{D_o}},\\
	\frac{\pd\tilde{I}_r}{\pd q}&=-\frac{\tilde{I}_r}{q}+\frac{2}{q\pa{4-q^2}}\br{\pa{8-q^2}\frac{\bar{R}}{D_s}+\frac{4\lambda}{D_s}},\\
	\frac{\pd J_r}{\pd\lambda}&=-\frac{1}{2\lambda}\pa{\frac{4\bar{R}+\lambda}{D_s}-\frac{7}{q}+\frac{3}{4}\frac{D_s-q\bar{R}}{\lambda}},\\
	\frac{\pd\tilde{J}_r}{\pd\lambda}&=\frac{1}{\lambda}\pa{\frac{4\bar{R}+\lambda}{D_s}+\frac{3}{4}\frac{D_s}{\lambda}},\\
	\frac{\pd J_r}{\pd q}&=\frac{7}{2q}I_r+\frac{1}{2}+\frac{3}{8q}\frac{D_s-q\bar{R}}{\lambda}-\frac{11}{q^2}\\
	&\quad+\frac{1}{2q\pa{4-q^2}}\br{\frac{\pa{32-5q^2}\bar{R}+\pa{16-q^2}\lambda}{D_s}+\frac{\pa{7-q^2}\pa{8-q^2+2R_o}}{D_o}-\frac{24}{q}},\nonumber\\
	\frac{\pd\tilde{J}_r}{\pd q}&=\frac{7}{2q}\tilde{I}_r-\frac{1}{q\pa{4-q^2}}\frac{\pa{32-5q^2}\bar{R}+\pa{16-q^2}\lambda}{D_s}-\frac{3}{4q}\frac{D_s}{\lambda}.
\end{align}
\end{subequations}

\section{Angular integrals}
\label{app:AngularIntegrals}

We quote relevant results from Ref.~\cite{InProgress}.  In terms of a new variable $u=\cos^2{\theta}$, we find that the angular potential $\Theta(u)=\hat{q}^2+u\br{a^2-\hat{\lambda}^2\pa{1-u}^{-1}}$ has roots at
\begin{align}
	u_\pm=\Delta_\theta\pm\sqrt{\Delta_\theta^2+\frac{\hat{q}^2}{a^2}},\qquad
	\Delta_\theta=\frac{1}{2}\pa{1-\frac{\hat{q}^2+\hat{\lambda}^2}{a^2}}.
\end{align}
As aforementioned in Eq.~\eqref{eq:AngularTurningPoints}, this implies that $\Theta(\theta)$ has roots at $\theta_\pm=\arccos{\mp\sqrt{u_+}}$, with the ordering $0<\theta_-<\theta_+<\pi$.  Note that we always have $u_-<0<u_+$.

The incomplete elliptic integrals of the first, second, and third kind are respectively defined as
\begin{subequations}
\begin{align}
	F(\phi|m)&=\int_0^\phi\frac{\ed\theta}{\sqrt{1-m\sin^2{\theta}}}
	=\int_0^{\sin{\phi}}\frac{\ed t}{\sqrt{\pa{1-t^2}\pa{1-mt^2}}},\\
	E(\phi|m)&=\int_0^\phi\sqrt{1-m\sin^2{\theta}}\ed\theta
	=\int_0^{\sin{\phi}}\sqrt{\frac{1-mt^2}{1-t^2}}\ed t,\\
	\Pi(n;\phi|m)&=\int_0^\phi\frac{1}{\pa{1-n\sin^2{\theta}}}\frac{\ed\theta}{\sqrt{1-m\sin^2{\theta}}}
	=\int_0^{\sin{\phi}}\frac{1}{1-nt^2}\frac{\ed t}{\sqrt{\pa{1-t^2}\pa{1-mt^2}}}.
\end{align}
\end{subequations}
Elliptic integrals are said to be ``complete'' when the amplitude $\phi=\pi/2$. The complete elliptic integrals of the first, second, and third kind may thus be resepectively defined as
\begin{subequations}
\begin{align}
	K(m)&=F\pa{\frac{\pi}{2}\Big|m}
	=\frac{\pi}{2}\,{_2F_1}\pa{\frac{1}{2},\frac{1}{2};1;m},\\
	E(m)&=E\pa{\frac{\pi}{2}\Big|m}
	=\frac{\pi}{2}\,{_2F_1}\pa{\frac{1}{2},-\frac{1}{2};1;m},\\
	\Pi(n|m)&=\Pi\pa{n;\frac{\pi}{2}\Big|m}.
\end{align}
\end{subequations}
where ${_2F_1}$ denotes Gauss' hypergeometric function.

The angular elliptic integrals that appear in the Kerr lens equations~\eqref{eq:LensEquations} are
\begin{subequations}
\begin{align}
	G_\theta&=M\int_{\theta_-}^{\theta_+}\frac{\ed\theta}{\sqrt{\Theta(\theta)}}
	=\frac{2M}{\ab{a}\sqrt{-u_-}}K\pa{\frac{u_+}{u_-}},\\
	\hat{G}_\theta&=M\int_{\theta_o}^{\pi/2}\frac{\ed\theta}{\sqrt{\Theta(\theta)}}
	=\frac{M}{\ab{a}\sqrt{-u_-}}F\pa{\Psi_o\left|\frac{u_+}{u_-}\right.},\\
	G_\phi&=M\int_{\theta_-}^{\theta_+}\frac{\csc^2{\theta}}{\sqrt{\Theta(\theta)}}\ed\theta
	=\frac{2M}{\ab{a}\sqrt{-u_-}}\Pi\pa{u_+\left|\frac{u_+}{u_-}\right.},\\
	\hat{G}_\phi&=M\int_{\theta_o}^{\pi/2}\frac{\csc^2{\theta}}{\sqrt{\Theta(\theta)}}\ed\theta
	=\frac{M}{\ab{a}\sqrt{-u_-}}\Pi\pa{u_+;\Psi_o\left|\frac{u_+}{u_-}\right.},\\
	G_t&=M\int_{\theta_-}^{\theta_+}\frac{\cos^2{\theta}}{\sqrt{\Theta(\theta)}}\ed\theta
	=-\frac{4Mu_+}{\ab{a}\sqrt{-u_-}}E'\pa{\frac{u_+}{u_-}},\\
	\hat{G}_t&=M\int_{\theta_o}^{\pi/2}\frac{\cos^2{\theta}}{\sqrt{\Theta(\theta)}}\ed\theta
	=-\frac{2Mu_+}{\ab{a}\sqrt{-u_-}}E'\pa{\Psi_o\left|\frac{u_+}{u_-}\right.},
\end{align}
\end{subequations}
where $E'(\phi|m)=\pd_mE(\phi|m)$ and
\begin{align}
	\Psi_o=\arcsin{\sqrt{\frac{\cos^2{\theta_o}}{u_+}}}.
\end{align}

In the near-extremal regime \eqref{eq:NearExtremalExpansion},
\begin{align}
	u_\pm=\mathcal{Q}_\pm+\O{\epsilon},\qquad
	\mathcal{Q}_\pm=\frac{q^2}{2}-3\pm\sqrt{12-\pa{2q}^2+\pa{\frac{q^2}{2}}^2}\gtrless0,\qquad
	\Psi_o=\arcsin{\sqrt{\frac{\cos^2{\theta_o}}{\mathcal{Q}_+}}}+\O{\epsilon}.
\end{align}
As such,
\begin{subequations}
\begin{align}
	G_\theta&=\frac{\pi}{\sqrt{-\mathcal{Q}_-}}{\,_2F_1}\pa{\frac{1}{2},\frac{1}{2},1,\frac{\mathcal{Q}_+}{\mathcal{Q}_-}}+\O{\epsilon},&
	\hat{G}_\theta&=\frac{1}{\sqrt{-\mathcal{Q}_-}}F\pa{\Psi_o\left|\frac{\mathcal{Q}_+}{\mathcal{Q}_+}\right.}+\O{\epsilon},\\
	G_\phi&=\frac{2}{\sqrt{-\mathcal{Q}_-}}\Pi\pa{\mathcal{Q}_+\left|\frac{\mathcal{Q}_+}{\mathcal{Q}_-}\right.}+\O{\epsilon},&
	\hat{G}_\phi&=\frac{1}{\sqrt{-\mathcal{Q}_-}}\Pi\pa{\mathcal{Q}_+;\Psi_o\left|\frac{\mathcal{Q}_+}{\mathcal{Q}_-}\right.}+\O{\epsilon},\\
	G_t&=-\frac{4\mathcal{Q}_+}{\sqrt{-\mathcal{Q}_-}}E'\pa{\frac{\mathcal{Q}_+}{\mathcal{Q}_-}}+\O{\epsilon},&
	\hat{G}_t&=-\frac{2\mathcal{Q}_+}{\sqrt{-\mathcal{Q}_-}}E'\pa{\Psi_o\left|\frac{\mathcal{Q}_+}{\mathcal{Q}_-}\right.}+\O{\epsilon}.
\end{align}
\end{subequations}
Now define
\begin{subequations}
\begin{align}
	G_{t\phi}&=2G_\phi-\frac{1}{2}G_t
	=\frac{2}{\sqrt{-\mathcal{Q}_-}}\br{2\Pi\pa{\mathcal{Q}_+\left|\frac{\mathcal{Q}_+}{\mathcal{Q}_-}\right.}+\mathcal{Q}_+E'\pa{\frac{\mathcal{Q}_+}{\mathcal{Q}_-}}}+\O{\epsilon},\\
	\hat{G}_{t\phi}&=2\hat{G}_\phi-\frac{1}{2}\hat{G}_t
	=\frac{1}{\sqrt{-\mathcal{Q}_-}}\br{2\Pi\pa{\mathcal{Q}_+;\Psi_o\left|\frac{\mathcal{Q}_+}{\mathcal{Q}_-}\right.}+\mathcal{Q}_+E'\pa{\Psi_o\left|\frac{\mathcal{Q}_+}{\mathcal{Q}_-}\right.}}+\O{\epsilon}.
\end{align}
\end{subequations}
Then in the near-extremal regime, we find that the $G$ integral defined in Eq.~\eqref{eq:RelevantG} becomes
\begin{align}
	G_{t\phi}^{m,s}=\frac{\hat{\lambda}G_\phi^{m,s}-\Omega_sa^2G_t^{m,s}}{M}
	=2G_\phi^{m,s}-\frac{1}{2}G_t^{m,s}+\O{\epsilon}
	=mG_{t\phi}-s\hat{G}_{t\phi}+\O{\epsilon}.
\end{align}

Finally, note that in our computation, we only need the variations of these expressions with respect to $q$, for which (at leading order in $\epsilon$), it suffices to take a simple derivative with respect to $q$.

\section{Screen coordinates}
\label{app:ScreenCoordinates}

\begin{figure}[ht!]
	\centering
	\includegraphics[width=.45\textwidth]{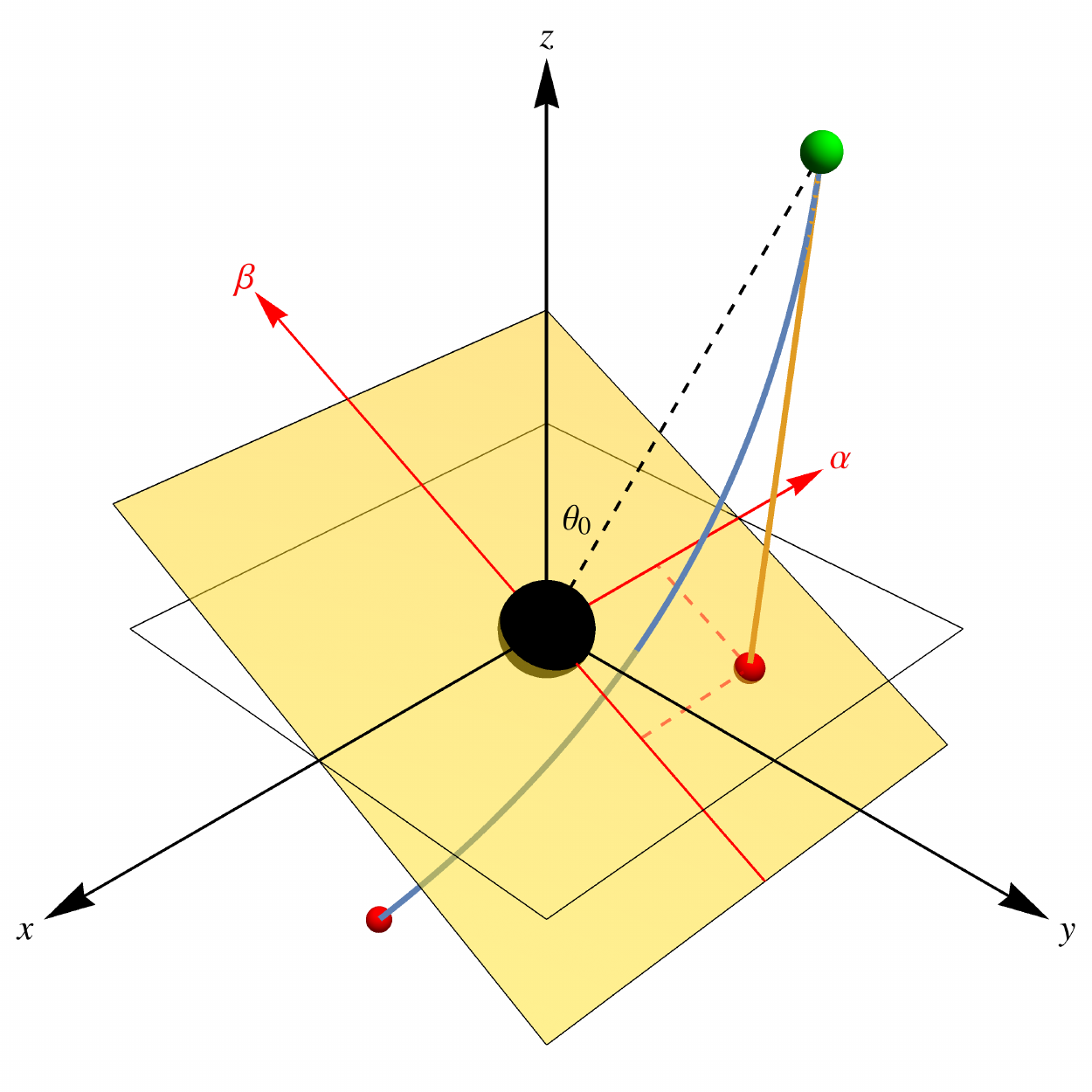}
	\caption{The screen coordinates $\pa{\alpha,\beta}$ are Cartesian coordinates for the apparent position on the plane of the sky.}
	\label{fig:ScreenCoordinates}
\end{figure}

In this section, we derive Eq.~\eqref{eq:ScreenCoordinates} for the screen coordinates $\pa{\alpha,\beta}$, which are just Cartesian coordinates for the apparent position on the plane of the sky.  The observer (green dot in Fig.~\ref{fig:ScreenCoordinates}) is assumed to be far away from the black hole, where the curvature is negligible and the spacetime is approximately flat.  The unit vectors are related to a Cartesian coordinate system $\cu{x,y,z}$ centered at the apparent position of the black hole by
\begin{align}
    \hat{\alpha}=-\sin{\phi_o}*\hat{x}+\cos{\phi_o}*\hat{y},\qquad
    \hat{\beta}=\mp\cos{\theta_o}\pa{\cos{\phi_o}*\hat{x}+\sin{\phi_o}*\hat{y}}\pm\sin{\theta_o}*\hat{z},
\end{align}
where the upper/lower sign corresponds to an image that lies in the northern/southern hemisphere of the observer's sky.  Let $\vec{x}_s$ denote the apparent position of the source on the screen (red dot in Fig.~\ref{fig:ScreenCoordinates}).  Then on the one hand,
\begin{align}
	\label{eq:ImagePosition}
	\vec{x}_s=\alpha*\hat{\alpha}+\beta*\hat{\beta}
	=-\pa{\alpha\sin{\phi_o}\pm\beta\cos{\theta_o}\cos{\phi_o}}\hat{x}+\pa{\alpha\cos{\phi_o}\mp\beta\cos{\theta_o}\sin{\phi_o}}\hat{y}\pm\beta\sin{\theta_o}*\hat{z}.
\end{align}
The apparent position $\vec{x}_s$ of the source is obtained by extending the unit tangent vector $\vec{v}_o$ of the of the photon at the observer a distance $r_o$ in the fictitious global Cartesian coordinate system (orange line in Fig.~\ref{fig:ScreenCoordinates}).  The unit tangent is given by $v_o^i=p_o^i/p_o^t$, where $i\in\cu{1,2,3}$ and $p_o^\mu$ is the four-momentum of the photon at the observer, expressed in Cartesian (asymptotically Minkowski) coordinates.  That is,
\begin{align}
	x^i_s=x^i_o-r_o\frac{p^i_o}{p_o^t}.
\end{align}
In terms of spherical coordinates, we find
\begin{align}
	\frac{\vec{x}_s}{r_o^2}=-\pa{\cos{\theta_o}\cos{\phi_o}\frac{p_o^\theta}{p_o^t}-\sin{\theta_o}\sin{\phi_o}\frac{p_o^\phi}{p_o^t}}\hat{x}-\pa{\cos{\theta_o}\sin{\phi_o}\frac{p_o^\theta}{p_o^t}+\sin{\theta_o}\cos{\phi_o}\frac{p_o^\phi}{p_o^t}}\hat{y}+\sin{\theta_o}\frac{p_o^\theta}{p_o^t}\hat{z}.
\end{align}
Equating the components of this expression to those in Eq.~\eqref{eq:ImagePosition} yields
\begin{align}
	\alpha=-r_o^2\sin{\theta_o}\frac{p_o^\phi}{p_o^t},\qquad
	\beta=\pm r_o^2\frac{p_o^\theta}{p_o^t}.
\end{align}
Using Eqs.~\eqref{eq:GeodesicEquation}, we find that as $r_o\to\infty$,
\begin{align}
	\pa{\alpha,\beta}&=\pa{-\frac{\hat{\lambda}}{\sin{\theta_o}},\pm\sqrt{\hat{q}^2+a^2\cos^2{\theta_o}-\hat{\lambda}^2\cot^2{\theta_o}}}+\O{\frac{1}{r_o}}.
\end{align}

\bibliographystyle{utphys}
\bibliography{NHEKlace}

\providecommand{\href}[2]{#2}\begingroup\raggedright\begin{thebibliography}{10}

\bibitem{Narayan2013}
R.~{Narayan} and J.~E. {McClintock}, ``{Observational Evidence for Black
  Holes},'' {\em ArXiv e-prints} (Dec., 2013) ,
  \href{http://arxiv.org/abs/1312.6698}{{\ttfamily arXiv:1312.6698
  [astro-ph.HE]}}.

\bibitem{Abbott2016a}
B.~P. {Abbott}, R.~{Abbott}, T.~D. {Abbott}, M.~R. {Abernathy}, F.~{Acernese},
  K.~{Ackley}, C.~{Adams}, T.~{Adams}, P.~{Addesso}, R.~X. {Adhikari}, {\em
  et~al.}, ``{Observation of Gravitational Waves from a Binary Black Hole
  Merger},'' \href{http://dx.doi.org/10.1103/PhysRevLett.116.061102}{{\em \prl}
  {\bfseries 116} no.~6, (Feb., 2016) 061102},
  \href{http://arxiv.org/abs/1602.03837}{{\ttfamily arXiv:1602.03837 [gr-qc]}}.

\bibitem{Abbott2016b}
B.~P. {Abbott}, R.~{Abbott}, T.~D. {Abbott}, M.~R. {Abernathy}, F.~{Acernese},
  K.~{Ackley}, C.~{Adams}, T.~{Adams}, P.~{Addesso}, R.~X. {Adhikari}, {\em
  et~al.}, ``{GW151226: Observation of Gravitational Waves from a 22-Solar-Mass
  Binary Black Hole Coalescence},''
  \href{http://dx.doi.org/10.1103/PhysRevLett.116.241103}{{\em \prl} {\bfseries
  116} no.~24, (June, 2016) 241103},
  \href{http://arxiv.org/abs/1606.04855}{{\ttfamily arXiv:1606.04855 [gr-qc]}}.

\bibitem{Abbott2017a}
B.~P. {Abbott}, R.~{Abbott}, T.~D. {Abbott}, F.~{Acernese}, K.~{Ackley},
  C.~{Adams}, T.~{Adams}, P.~{Addesso}, R.~X. {Adhikari}, V.~B. {Adya}, {\em
  et~al.}, ``{GW170104: Observation of a 50-Solar-Mass Binary Black Hole
  Coalescence at Redshift 0.2},''
  \href{http://dx.doi.org/10.1103/PhysRevLett.118.221101}{{\em \prl} {\bfseries
  118} no.~22, (June, 2017) 221101},
  \href{http://arxiv.org/abs/1706.01812}{{\ttfamily arXiv:1706.01812 [gr-qc]}}.

\bibitem{Abbott2017b}
B.~P. {Abbott}, R.~{Abbott}, T.~D. {Abbott}, F.~{Acernese}, K.~{Ackley},
  C.~{Adams}, T.~{Adams}, P.~{Addesso}, R.~X. {Adhikari}, V.~B. {Adya}, {\em
  et~al.}, ``{GW170814: A Three-Detector Observation of Gravitational Waves
  from a Binary Black Hole Coalescence},''
  \href{http://dx.doi.org/10.1103/PhysRevLett.119.141101}{{\em \prl} {\bfseries
  119} no.~14, (Oct., 2017) 141101},
  \href{http://arxiv.org/abs/1709.09660}{{\ttfamily arXiv:1709.09660 [gr-qc]}}.

\bibitem{Bardeen1973}
J.~M. {Bardeen}, ``{Timelike and null geodesics in the Kerr metric},'' in {\em
  Black Holes (Les Astres Occlus)}, C.~{Dewitt} and B.~S. {Dewitt}, eds.,
  pp.~215--239.
\newblock Gordon and Breach, New York, NY, 1973.

\bibitem{Luminet1979}
J.-P. {Luminet}, ``{Image of a spherical black hole with thin accretion
  disk},'' {\em \aap} {\bfseries 75} (May, 1979) 228--235.

\bibitem{Broderick2006}
A.~E. {Broderick} and A.~{Loeb}, ``{Imaging optically-thin hotspots near the
  black hole horizon of Sgr A* at radio and near-infrared wavelengths},''
  \href{http://dx.doi.org/10.1111/j.1365-2966.2006.10152.x}{{\em \mnras}
  {\bfseries 367} (Apr., 2006) 905--916},
  \href{http://arxiv.org/abs/astro-ph/0509237}{{\ttfamily astro-ph/0509237}}.

\bibitem{Doeleman2008a}
S.~S. {Doeleman}, J.~{Weintroub}, A.~E.~E. {Rogers}, R.~{Plambeck},
  R.~{Freund}, R.~P.~J. {Tilanus}, P.~{Friberg}, L.~M. {Ziurys}, J.~M. {Moran},
  B.~{Corey}, K.~H. {Young}, D.~L. {Smythe}, M.~{Titus}, D.~P. {Marrone}, R.~J.
  {Cappallo}, D.~C.-J. {Bock}, G.~C. {Bower}, R.~{Chamberlin}, G.~R. {Davis},
  T.~P. {Krichbaum}, J.~{Lamb}, H.~{Maness}, A.~E. {Niell}, A.~{Roy},
  P.~{Strittmatter}, D.~{Werthimer}, A.~R. {Whitney}, and D.~{Woody},
  ``{Event-horizon-scale structure in the supermassive black hole candidate at
  the Galactic Centre},'' \href{http://dx.doi.org/10.1038/nature07245}{{\em
  \nat} {\bfseries 455} (Sept., 2008) 78--80},
  \href{http://arxiv.org/abs/0809.2442}{{\ttfamily arXiv:0809.2442}}.

\bibitem{Doeleman2008b}
S.~S. {Doeleman}, V.~L. {Fish}, A.~E. {Broderick}, A.~{Loeb}, and A.~E.~E.
  {Rogers}, ``{Detecting Flaring Structures in Sagittarius A* with
  High-Frequency VLBI},''
  \href{http://dx.doi.org/10.1088/0004-637X/695/1/59}{{\em \apj} {\bfseries
  695} (Apr., 2009) 59--74}, \href{http://arxiv.org/abs/0809.3424}{{\ttfamily
  arXiv:0809.3424}}.

\bibitem{Doeleman2009}
S.~{Doeleman}, E.~{Agol}, D.~{Backer}, F.~{Baganoff}, G.~C. {Bower},
  A.~{Broderick}, A.~{Fabian}, V.~{Fish}, C.~{Gammie}, P.~{Ho}, M.~{Honman},
  T.~{Krichbaum}, A.~{Loeb}, D.~{Marrone}, M.~{Reid}, A.~{Rogers},
  I.~{Shapiro}, P.~{Strittmatter}, R.~{Tilanus}, J.~{Weintroub}, A.~{Whitney},
  M.~{Wright}, and L.~{Ziurys}, ``{Imaging an Event Horizon: submm-VLBI of a
  Super Massive Black Hole},'' in {\em astro2010: The Astronomy and
  Astrophysics Decadal Survey}, vol.~2010 of {\em Astronomy}.
\newblock 2009.
\newblock \href{http://arxiv.org/abs/0906.3899}{{\ttfamily arXiv:0906.3899
  [astro-ph.CO]}}.

\bibitem{Doeleman2012}
S.~S. {Doeleman}, V.~L. {Fish}, D.~E. {Schenck}, C.~{Beaudoin}, R.~{Blundell},
  G.~C. {Bower}, A.~E. {Broderick}, R.~{Chamberlin}, R.~{Freund}, P.~{Friberg},
  M.~A. {Gurwell}, P.~T.~P. {Ho}, M.~{Honma}, M.~{Inoue}, T.~P. {Krichbaum},
  J.~{Lamb}, A.~{Loeb}, C.~{Lonsdale}, D.~P. {Marrone}, J.~M. {Moran},
  T.~{Oyama}, R.~{Plambeck}, R.~A. {Primiani}, A.~E.~E. {Rogers}, D.~L.
  {Smythe}, J.~{SooHoo}, P.~{Strittmatter}, R.~P.~J. {Tilanus}, M.~{Titus},
  J.~{Weintroub}, M.~{Wright}, K.~H. {Young}, and L.~M. {Ziurys},
  ``{Jet-Launching Structure Resolved Near the Supermassive Black Hole in
  M87},'' \href{http://dx.doi.org/10.1126/science.1224768}{{\em Science}
  {\bfseries 338} (Oct., 2012) 355},
  \href{http://arxiv.org/abs/1210.6132}{{\ttfamily arXiv:1210.6132
  [astro-ph.HE]}}.

\bibitem{Johnson2015}
M.~D. {Johnson}, V.~L. {Fish}, S.~S. {Doeleman}, D.~P. {Marrone}, R.~L.
  {Plambeck}, J.~F.~C. {Wardle}, K.~{Akiyama}, K.~{Asada}, C.~{Beaudoin},
  L.~{Blackburn}, R.~{Blundell}, G.~C. {Bower}, C.~{Brinkerink}, A.~E.
  {Broderick}, R.~{Cappallo}, A.~A. {Chael}, G.~B. {Crew}, J.~{Dexter},
  M.~{Dexter}, R.~{Freund}, P.~{Friberg}, R.~{Gold}, M.~A. {Gurwell}, P.~T.~P.
  {Ho}, M.~{Honma}, M.~{Inoue}, M.~{Kosowsky}, T.~P. {Krichbaum}, J.~{Lamb},
  A.~{Loeb}, R.-S. {Lu}, D.~{MacMahon}, J.~C. {McKinney}, J.~M. {Moran},
  R.~{Narayan}, R.~A. {Primiani}, D.~{Psaltis}, A.~E.~E. {Rogers},
  K.~{Rosenfeld}, J.~{SooHoo}, R.~P.~J. {Tilanus}, M.~{Titus},
  L.~{Vertatschitsch}, J.~{Weintroub}, M.~{Wright}, K.~H. {Young}, J.~A.
  {Zensus}, and L.~M. {Ziurys}, ``{Resolved magnetic-field structure and
  variability near the event horizon of Sagittarius A*},''
  \href{http://dx.doi.org/10.1126/science.aac7087}{{\em Science} {\bfseries
  350} (Dec., 2015) 1242--1245},
  \href{http://arxiv.org/abs/1512.01220}{{\ttfamily arXiv:1512.01220
  [astro-ph.HE]}}.

\bibitem{Andersson2000}
N.~{Andersson} and K.~{Glampedakis}, ``{Superradiance Resonance Cavity Outside
  Rapidly Rotating Black Holes},''
  \href{http://dx.doi.org/10.1103/PhysRevLett.84.4537}{{\em Physical Review
  Letters} {\bfseries 84} (May, 2000) 4537--4540},
  \href{http://arxiv.org/abs/gr-qc/9909050}{{\ttfamily gr-qc/9909050}}.

\bibitem{Glampedakis2001}
K.~{Glampedakis} and N.~{Andersson}, ``{Late-time dynamics of rapidly rotating
  black holes},'' \href{http://dx.doi.org/10.1103/PhysRevD.64.104021}{{\em
  \prd} {\bfseries 64} no.~10, (Nov., 2001) 104021},
  \href{http://arxiv.org/abs/gr-qc/0103054}{{\ttfamily gr-qc/0103054}}.

\bibitem{Yang2013}
H.~{Yang}, A.~{Zimmerman}, A.~{Zengino{\u g}lu}, F.~{Zhang}, E.~{Berti}, and
  Y.~{Chen}, ``{Quasinormal modes of nearly extremal Kerr spacetimes: Spectrum
  bifurcation and power-law ringdown},''
  \href{http://dx.doi.org/10.1103/PhysRevD.88.044047}{{\em \prd} {\bfseries 88}
  no.~4, (Aug., 2013) 044047}, \href{http://arxiv.org/abs/1307.8086}{{\ttfamily
  arXiv:1307.8086 [gr-qc]}}.

\bibitem{Porfyriadis2014a}
A.~P. {Porfyriadis} and A.~{Strominger}, ``{Gravity waves from the Kerr/CFT
  correspondence},'' \href{http://dx.doi.org/10.1103/PhysRevD.90.044038}{{\em
  \prd} {\bfseries 90} no.~4, (Aug., 2014) 044038},
  \href{http://arxiv.org/abs/1401.3746}{{\ttfamily arXiv:1401.3746 [hep-th]}}.

\bibitem{Porfyriadis2014b}
S.~{Hadar}, A.~P. {Porfyriadis}, and A.~{Strominger}, ``{Gravity waves from
  extreme-mass-ratio plunges into Kerr black holes},''
  \href{http://dx.doi.org/10.1103/PhysRevD.90.064045}{{\em \prd} {\bfseries 90}
  no.~6, (Sept., 2014) 064045},
  \href{http://arxiv.org/abs/1403.2797}{{\ttfamily arXiv:1403.2797 [hep-th]}}.

\bibitem{Hadar2015}
S.~{Hadar}, A.~P. {Porfyriadis}, and A.~{Strominger}, ``{Fast plunges into Kerr
  black holes},'' \href{http://dx.doi.org/10.1007/JHEP07(2015)078}{{\em Journal
  of High Energy Physics} {\bfseries 7} (July, 2015) 78},
  \href{http://arxiv.org/abs/1504.07650}{{\ttfamily arXiv:1504.07650
  [hep-th]}}.

\bibitem{Gralla2016a}
S.~E. {Gralla}, A.~{Lupsasca}, and A.~{Strominger}, ``{Near-horizon Kerr
  magnetosphere},'' \href{http://dx.doi.org/10.1103/PhysRevD.93.104041}{{\em
  \prd} {\bfseries 93} no.~10, (May, 2016) 104041},
  \href{http://arxiv.org/abs/1602.01833}{{\ttfamily arXiv:1602.01833
  [hep-th]}}.

\bibitem{Gralla2016b}
S.~E. {Gralla}, S.~A. {Hughes}, and N.~{Warburton}, ``{Inspiral into
  Gargantua},'' \href{http://dx.doi.org/10.1088/0264-9381/33/15/155002}{{\em
  Classical and Quantum Gravity} {\bfseries 33} no.~15, (Aug., 2016) 155002},
  \href{http://arxiv.org/abs/1603.01221}{{\ttfamily arXiv:1603.01221 [gr-qc]}}.

\bibitem{Burko2016}
L.~M. {Burko} and G.~{Khanna}, ``{Gravitational waves from a plunge into a
  nearly extremal Kerr black hole},''
  \href{http://dx.doi.org/10.1103/PhysRevD.94.084049}{{\em \prd} {\bfseries 94}
  no.~8, (Oct., 2016) 084049},
  \href{http://arxiv.org/abs/1608.02244}{{\ttfamily arXiv:1608.02244 [gr-qc]}}.

\bibitem{Hadar2017}
S.~{Hadar} and A.~P. {Porfyriadis}, ``{Whirling orbits around twirling black
  holes from conformal symmetry},''
  \href{http://dx.doi.org/10.1007/JHEP03(2017)014}{{\em Journal of High Energy
  Physics} {\bfseries 3} (Mar., 2017) 14},
  \href{http://arxiv.org/abs/1611.09834}{{\ttfamily arXiv:1611.09834
  [hep-th]}}.

\bibitem{Compere2017}
G.~{Comp{\`e}re} and R.~{Oliveri}, ``{Self-similar accretion in thin discs
  around near-extremal black holes},''
  \href{http://dx.doi.org/10.1093/mnras/stx748}{{\em \mnras} {\bfseries 468}
  (July, 2017) 4351--4361}, \href{http://arxiv.org/abs/1703.00022}{{\ttfamily
  arXiv:1703.00022 [astro-ph.HE]}}.

\bibitem{Bardeen1999}
J.~{Bardeen} and G.~T. {Horowitz}, ``{Extreme Kerr throat geometry: A vacuum
  analog of AdS\_2 x S\^2},''
  \href{http://dx.doi.org/10.1103/PhysRevD.60.104030}{{\em \prd} {\bfseries 60}
  no.~10, (Nov., 1999) 104030},
  \href{http://arxiv.org/abs/hep-th/9905099}{{\ttfamily hep-th/9905099}}.

\bibitem{Guica2009}
M.~{Guica}, T.~{Hartman}, W.~{Song}, and A.~{Strominger}, ``{The Kerr/CFT
  correspondence},'' \href{http://dx.doi.org/10.1103/PhysRevD.80.124008}{{\em
  \prd} {\bfseries 80} no.~12, (Dec., 2009) 124008},
  \href{http://arxiv.org/abs/0809.4266}{{\ttfamily arXiv:0809.4266 [hep-th]}}.

\bibitem{Bardeen1972}
J.~M. {Bardeen}, W.~H. {Press}, and S.~A. {Teukolsky}, ``{Rotating Black Holes:
  Locally Nonrotating Frames, Energy Extraction, and Scalar Synchrotron
  Radiation},'' \href{http://dx.doi.org/10.1086/151796}{{\em \apj} {\bfseries
  178} (Dec., 1972) 347--370}.

\bibitem{Cunningham1972}
C.~T. {Cunningham} and J.~M. {Bardeen}, ``{The Optical Appearance of a Star
  Orbiting an Extreme Kerr Black Hole},''
  \href{http://dx.doi.org/10.1086/180933}{{\em \apjl} {\bfseries 173} (May,
  1972) L137}.

\bibitem{Cunningham1973}
C.~T. {Cunningham} and J.~M. {Bardeen}, ``{The Optical Appearance of a Star
  Orbiting an Extreme Kerr Black Hole},''
  \href{http://dx.doi.org/10.1086/152223}{{\em \apj} {\bfseries 183} (July,
  1973) 237--264}.

\bibitem{Vazquez2004}
S.~E. {V{\'a}zquez} and E.~P. {Esteban}, ``{Strong-field gravitational lensing
  by a Kerr black hole},''
  \href{http://dx.doi.org/10.1393/ncb/i2004-10121-y}{{\em Nuovo Cimento B
  Serie} {\bfseries 119} (May, 2004) 489},
  \href{http://arxiv.org/abs/gr-qc/0308023}{{\ttfamily gr-qc/0308023}}.

\bibitem{InProgress}
A.~{Lupsasca}, ``{Kerr angular geodesic integrals},'' in {\em In preparation}.
\newblock 2017.

\bibitem{Johnson1982}
M.~H. {Johnson} and E.~{Teller}, ``{Intensity Changes in the Doppler Effect},''
  {\em Proceedings of the National Academy of Science} {\bfseries 79} (Feb.,
  1982) 1340--1340.

\bibitem{Porfyriadis2016}
A.~P. {Porfyriadis}, Y.~{Shi}, and A.~{Strominger}, ``{Photon emission near
  extreme Kerr black holes},''
  \href{http://dx.doi.org/10.1103/PhysRevD.95.064009}{{\em \prd} {\bfseries 95}
  no.~6, (Mar., 2017) 064009},
  \href{http://arxiv.org/abs/1607.06028}{{\ttfamily arXiv:1607.06028 [gr-qc]}}.

\bibitem{Gralla2015}
S.~E. {Gralla}, A.~P. {Porfyriadis}, and N.~{Warburton}, ``{Particle on the
  innermost stable circular orbit of a rapidly spinning black hole},''
  \href{http://dx.doi.org/10.1103/PhysRevD.92.064029}{{\em \prd} {\bfseries 92}
  no.~6, (Sept., 2015) 064029},
  \href{http://arxiv.org/abs/1506.08496}{{\ttfamily arXiv:1506.08496 [gr-qc]}}.

\bibitem{Carlson2015}
E.~{Carlson}, T.~{Jeltema}, and S.~{Profumo}, ``{Where do the 3.5 keV photons
  come from? A morphological study of the Galactic Center and of Perseus},''
  \href{http://dx.doi.org/10.1088/1475-7516/2015/02/009}{{\em \jcap} {\bfseries
  2} (Feb., 2015) 009}, \href{http://arxiv.org/abs/1411.1758}{{\ttfamily
  arXiv:1411.1758 [astro-ph.HE]}}.

\bibitem{Gralla2016c}
S.~E. {Gralla}, A.~{Zimmerman}, and P.~{Zimmerman}, ``{Transient instability of
  rapidly rotating black holes},''
  \href{http://dx.doi.org/10.1103/PhysRevD.94.084017}{{\em \prd} {\bfseries 94}
  no.~8, (Oct., 2016) 084017},
  \href{http://arxiv.org/abs/1608.04739}{{\ttfamily arXiv:1608.04739 [gr-qc]}}.

\bibitem{Abramowicz2013}
M.~A. {Abramowicz} and P.~C. {Fragile}, ``{Foundations of Black Hole Accretion
  Disk Theory},'' \href{http://dx.doi.org/10.12942/lrr-2013-1}{{\em Living
  Reviews in Relativity} {\bfseries 16} (Jan., 2013) 1},
  \href{http://arxiv.org/abs/1104.5499}{{\ttfamily arXiv:1104.5499
  [astro-ph.HE]}}.

\end{thebibliography}\endgroup

\end{document}